\documentclass[12pt]{article}

\usepackage{amsmath,amsfonts,amsthm,amssymb}
\usepackage{eepic,fullpage,graphicx,color}

\usepackage[tight]{subfigure}

\newtheorem{proposition}{Proposition}[section]

\newtheorem{theorem}{Theorem}
\newtheorem{lemma}{Lemma}[section]
\newtheorem{assumption}{Assumption}

\newcommand{\cov}{{\rm Cov}}
\newcommand{\var}{{\rm Var}}
\newcommand{\tl}{T_L}
\newcommand{\tr}{T_R}
\newcommand{\rl}{\rho_L}
\newcommand{\rr}{\rho_R}

\newtheorem{nexp}{Simulation}

\newcommand{\putgraph}[2]{\includegraphics[#1]{#2}}

\newcommand{\phibc}{\varphi^{bc}}
\newcommand{\phihat}{\hat\varphi^{bc}}

\newcommand{\eps}{\varepsilon}

\newcommand{\kaption}[1]{\caption{\small #1}}

\title{\Large Correlations in Nonequilibrium Steady States\\
of Random Halves Models}

\author{Kevin K. Lin\footnote{E-mail: klin@cims.nyu.edu\ . This research 
was supported by an NSF Postdoctoral Fellowship.}\hspace{0.5in}
Lai-Sang Young\footnote{E-mail: lsy@cims.nyu.edu\ . This research
was supported by a grant from the NSF.}\vspace{6pt}\\
{\large Courant Institute of Mathematical Sciences}\\
{\large New York University}}

\date{\small February 23, 2007}

\begin{document}
\maketitle

\begin{abstract}
We present the results of a detailed study of energy correlations at
steady state for a 1-D model of coupled energy and matter transport.
Our aim is to discover --- via theoretical arguments, conjectures, and
numerical simulations --- how spatial covariances scale with system
size, their relations to local thermodynamic quantities, and the
randomizing effects of heat baths.  Among our findings are that
short-range covariances respond quadratically to local temperature
gradients, and long-range covariances decay linearly with macroscopic
distance.  These findings are consistent with exact results for the
simple exclusion and KMP models.
\end{abstract}

\begin{small}
\tableofcontents
\end{small}

\vskip .5in

\section*{Introduction}
\addcontentsline{toc}{section}{\numberline {}Introduction}

Transport processes, such as heat flow through a conducting medium in
contact with unequal heat reservoirs, are intrinsically nonequilibrium
phenomena because of the presence of nonzero currents~\cite{dm,lepri1}.
A central problem in nonequilibrium statistical physics is to explain
how such large-scale, macroscopic processes arise from complex
microscopic interactions.  The most basic questions are perhaps those of
mean profiles of quantities of physical interest and their responses to
external forces.  Nonequilibrium steady states, on the other hand, are
well known to be characterized by large fluctuations, and among the
simplest measures of fluctuations are temporal and spatial correlations.

In this paper, we present a systematic study of spatial correlations at
steady state for a class of 1-D stochastic models called Random Halves
Models.  A detailed description of this class of models is given in
Sect.~\ref{sec:ey-desc}.  Very briefly, a linear chain of open cells is
connected to two unequal heat baths, which inject into the chain tracer
particles at characteristic rates and energies.  Energy storage devices
are systematically placed throughout the chain to mark local energy
levels.  Their contents are redistributed by the tracer particles as
they move through the chain.  These models were introduced in
{\cite{ey}} as stochastic idealizations of certain mechanical models
{\cite{llm,mll}}.  

Our study is based on a combination of analytic arguments and numerical
simulations.  We believe random halves models are excellent candidates
for this method of investigation for two reasons.  First, their dynamics
are richer and more complex than rigorously-understood models such as
the simple exclusion {\cite{derrida}} and KMP models {\cite{kipnis}}, as
random halves models have two transported quantities (energy and matter)
and highly nonlinear interactions involving two different time scales.
These features make a purely analytical study more difficult.  Second,
all of the forces acting on this system are clearly identified.  This is
seldom the case in more realistic physical models.

Our main results can be summarized as follows.  Given boundary
conditions, {\it i.e.}, the temperature and injection rate of each bath,
we let $S_i$ denote the stored energy at the $i$th site, and
$\cov_N(S_i,S_j)$ the covariance of $S_i$ and $S_j$ in a chain of length
$N$.  Our first finding is $\cov_N(S_i,S_j) \sim \frac{1}{N}$ for $i\neq
j$, which leads us to consider the following two functions
describing the covariances of stored energies at microscopic and
macroscopic distances:
\begin{eqnarray*}
{\cal C}(x) & = & \lim_{N \to \infty} N \cdot 
\cov_N(S_{[xN]},S_{[xN]+1})\ ,
\qquad x \in (0,1);\\
{\cal C}_2(x,y) & = & \lim_{N \to \infty} N \cdot \cov_N(S_{[xN]},S_{[yN]})\ ,
\qquad x,y \in (0,1),\ \ x\neq y.
\end{eqnarray*}
We show that ${\cal C}(x)$ has the form
\begin{equation}
{\cal C}(x) \ \approx \ \hat\varphi^{bc}(x) \cdot A(\kappa(x)) \cdot
T'(x)^2
\end{equation}
where the right side is to be interpreted as follows:
$\hat\varphi^{bc}(x)$ is a measure of effective distance to the baths, a
rough approximation of which is $\hat\varphi^{bc}(x) \approx 4x(1-x)$;
$T(x)$ is local temperature, and $T'(x)$ is temperature gradient;
$\kappa(x)$ is local particle density, and $A(\kappa) \to 0$ as $\kappa
\to \infty$. Note the quadratic dependence on the local temperature
gradient.  For long-range covariances, we find that for fixed $x$, the
function $y \mapsto {\cal C}_2(x,y)$ is continuous but not
differentiable at $y=x$, decreasing roughly linearly as $|y-x|$
increases.

Our numerical results are consistent with exact analytic results on the
simple exclusion and KMP models~\cite{derrida,derrida1,spohn,bertini2}.
We extend the existing picture to a setting of coupled energy and matter
transport, with features not present in these two previously studied
models.  For example, we show that energy covariances at
microscopic distances respond quadratically to local temperature
gradients (which vary along the chain; see~\cite{ey}).  There is a
second transported quantity, namely matter in the form of particles, and
energy covariances are shown to be inversely related to local particle
density.  Throughout the paper, we provide cross-checking numerical
evidence for the phenomena we identify, and venture to give a physical
interpretation whenever we can.
See~\cite{bertini1,bertini1a,garrido,kipnis-landim,kirkpatrick} for
other relevant works on this topic.

While most of the results presented here are specific to 1-D, they also
serve as a basis for direct comparisons with higher dimensions.  A
corresponding analysis of random halves models in 2 and 3-D is underway.
The results will be reported in a separate paper.

\paragraph{Note on simulations.}
All of our numerical results are obtained via direct
simulation.  That is, our computer programs faithfully
implement the dynamics of the random halves model described
in Sect.~\ref{sec:ey-desc}, and expectation values with respect
to the invariant measures are computed as time averages over
long trajectories.  Some relevant numerical issues are discussed in
Appendix A.

\section{The Random Halves Model}
\label{sec:ey}

\subsection{Model Description and Physical Interpretation}
\label{sec:ey-desc}

A class of models of nonequilibrium phenomena is introduced in
{\cite{ey}}. In each model, there is a homogeneous conducting medium
represented by a linear chain of $N$ identical ``cells'' with stochastic
heat baths coupled to the ends of the chain.  Within each cell, there is
a {\it stored energy} which characterizes the ``temperature'' at that
location.  Matter (in the form of tracer particles) and energy (in the
form of tracer kinetic energy) are injected into the system by the heat
baths; they are eventually absorbed by the heat baths.  Tracer particles
interact with the local system at each site, redistributing the stored
energies as they move about in the chain; they do not interact directly
with each other.  This general framework was introduced as an
abstraction of the mechanical models in {\cite{llm,mll}}.  It
encompasses both ``Hamiltonian'' models with conservative deterministic
dynamics (the heat baths being the only sources of randomness), and
stochastic models, in which the microscopic dynamics are defined by
conservative stochastic rules.

\begin{figure}
\begin{center}
\putgraph{bb=0 0 612 113,scale=0.75}{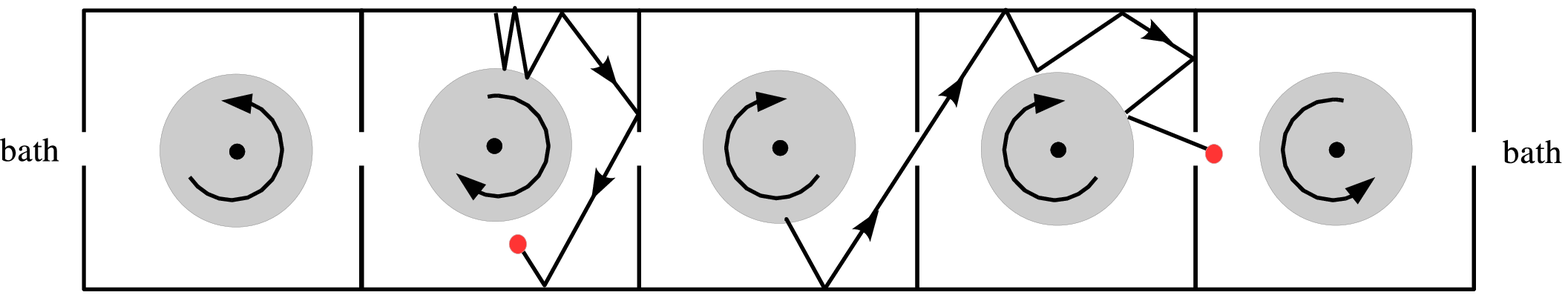}
\vspace{14pt}
\end{center}
\kaption{The rotating disc model.  The ``random halves'' model 
studied in this paper is a stochastic version of this model. }
\label{fig-model}
\end{figure}

For concreteness, we begin by describing a slightly simplified version
of the mechanical models studied in {\cite{llm,mll}}.  In this {\em
rotating disc model}, each cell contains a disc nailed down at its
center, about which it rotates freely (see Fig.~\ref{fig-model}).
Whenever a tracer collides with a disc, it exchanges kinetic energy with
the disc via a deterministic rule, {\em e.g.} the angular momentum of
the disc may be interchanged with the tangential component of the
tracer's momentum.  When a tracer collides with a cell wall, it reflects
elastically.  Here, the stored energy at each site is the rotational
energy of the disc.

The random halves model studied in the present paper are stochastic
idealizations of mechanical models like the rotating disc model
described above.  Nonequilibrium energy and particle density profiles of
random halves models have been analyzed in {\cite{ey}}.  The rest of
this section reviews relevant parts of that paper.

\paragraph{Precise description of the Random Halves Model.} 
There are $N$ sites, labeled $1,2,\cdots N$, with tracers moving through
the chain.  At each site there is an abstract {\em energy storage tank}.
We let $S_i(t)$ denote the amount of energy in the tank at site $i$ at
time $t$.  The microscopic dynamics are defined by the following rules.
Fix $\delta>0$.  Each tracer is equipped with two independent
exponential clocks.  Clock 1, which signals the times of energy
exchanges with tanks, rings at rate $\sqrt{e(t)}/\delta$, where $e$ is
the kinetic energy of the tracer.  Clock 2, which signals the times of
site-to-site movements, rings at rate $\sqrt{e(t)}$.  Suppose a tracer
is at site $i$ and one of its clocks rings.  Then instantaneously:
\begin{enumerate}

\item If Clock 1 rings, the tracer energy $e$ and the stored energy
$S_i$ are pooled together and split randomly.  That is, the tracer gets
$U\cdot(e+S_i)$ units of energy and the tank gets $(1-U)\cdot(e+S_i)$,
where $U \in [0,1]$ is uniformly distributed and independent of all
other random variables.

\item If Clock 2 rings, the tracer leaves site $i$.  It jumps with equal
probability to sites $i\pm 1$.  A tracer entering site $0$ or site $N+1$
exits the system forever.

\end{enumerate}
All tracers originate from and eventually exit to one of the heat baths.
Each heat bath injects tracers with independent,
exponentially-distributed energies into the system.  The left bath
injects tracers with mean energy $T_L$ into site 1 at an exponential
rate of $\rl$, and the right bath injects tracers with mean energy $T_R$
into site $N$ at a rate of $\rr$.  Tracers in the system are
indistinguishable.

\paragraph{Physical interpretation and remarks.}
It is natural to think of the parameters $\tl$ and $\tr$ as
temperatures.  This leads us to define the temperature $T_i$ at site $i$
in an $N$-chain to be $T_i := \mathbb{E}(S_i)$, where the expectation is
taken with respect to the invariant measure of the $N$-chain.  The
injection rates $\rho_L$ and $\rho_R$ can be rewritten in terms of
chemical potentials; we do not pursue this analogy further.

The Hamiltonian model contains a small length parameter which does not
appear in the stochastic model, namely a length $\ell$ which measures
the size of the cell.  In the stochastic model, we set $\ell=1$. It
is useful to keep this in mind in dimensional analysis.

There are two time scales in our system, one associated with the local
dynamics at each site and the other the movement of tracers along the
chain.  The ratio of these two time scales is $\delta$. For example,
$\delta\ll 1$ means that on average, tracer-tank energy exchanges occur
much more frequently than site-to-site movements of tracers.
Steady-state macroscopic profiles such as temperature and tracer density
do not depend on $\delta$, but $\delta$ can have a significant impact on
the numerical values of spatial and temporal correlations.  In this
paper, we have chosen to simplify matters by {\em fixing} $\delta$,
which throughout the paper is set equal to $\frac{1}{10}$.

At any moment in time, the number of tracers at each site can vary from
$0$ to $\infty$.  Observe also that the interaction in this simple model
is highly nonlinear, even though there are no direct tracer-tracer or
tank-tank interactions: tracers at the same site exchange energy via
interacting with the tank, and all actions --- including energy
exchanges and site-to-site jumps --- take place at rates proportional to
the ``speeds'' (the square roots of the kinetic energies) of the tracers
at that moment in time.  This is essential if these stochastic models
are to mimic the behaviors of their Hamiltonian counterparts.

\subsection{Invariant Measures at Equilibrium}
\label{sec:ey-equilibrium}

At equilibrium, {\it i.e.}, when the left and right baths have equal
temperatures and injection rates, the invariant measure of the random
halves model is known explicitly.  To give its density, we need a little
bit of notation: the state of a single cell in which there are exactly
$k$ tracers is specified by $(S, \{e_1,e_2,\cdots,e_k\})$ where $S > 0$
is the stored energy, {\it i.e.}, the energy of the tank, and
$\{e_1,e_2,\cdots,e_k\}$ is an unordered set of $k$ positive numbers
representing the $k$ tracer energies\footnote{It is an unordered set
because the tracers are indistinguishable.}.  Let $\Omega_k$ denote the
set of all possible states of a single cell with exactly $k$ tracers
present.  The state space for a single cell is then the disjoint union
$\Omega =\cup_k\Omega_k$.

\begin{proposition} {\cite{ey}}
\label{prop2.1}
Let $T_L =T_R =T$, and $\rho_L =\rho_R =\rho$. Then the 
unique invariant probability measure of the $N$-chain is the $N$-fold
product
$$
\mu_N \ =  \ \mu^{T,\rho} \times \cdots \times \mu^{T,\rho}
$$
where $\mu^{T,\rho}$ is the measure giving the statistics within each cell. The measure  $\mu^{T,\rho}$ is defined by:
\begin{enumerate}
\item The number of tracers present is a Poisson random variable with
mean $\kappa\equiv 2\rho \sqrt {\pi/ T}$, {\em i.e.},
\begin{equation}
\mu^{T,\rho}(\Omega_k) \ = \ p_k \ := \ \frac{\kappa^k}{k!} e^{-\kappa}~, 
\qquad k=0,1,2,\ldots.
\end{equation}
\item The conditional density of $\mu^{T,\rho}$ on $\Omega_k$ is
$c_k\sigma_k(\{e_1,\cdots,e_k\},S)$ where
\begin{equation}\label{(b)}
\sigma_k(\{e_1, \dots, e_k\}, S) \ = \ \frac{1}{\sqrt{e_1 \cdot\ldots\cdot e_k}} 
\ e^{-\beta (e_1+\dots + e_k +S)}~;
\end{equation}
here $\beta ={1}/{T}$, and $c_k=\beta \,k! \left(\beta/\pi\right)^{k/2}$ 
is the normalizing constant.
\end{enumerate}
\end{proposition}

Proposition {\ref{prop2.1}} shows that at equilibrium, the stored
energies at different sites are entirely uncorrelated. A generalization
of this result is given in Proposition 5.1.

\subsection{Macroscopic Equations of Nonequilibrium Steady States}
\label{sec:ey-eqs}

We now fix arbitrary boundary conditions $T_L$, $T_R$, 
$\rho_L$, and $\rho_R$,  and give the equations of temperature 
and tracer profiles in nonequilibrium steady states. 

\begin{assumption} We assume that for each $N$, there is
a unique invariant probability measure $\mu_N$ to which all initial
data converge.
\end{assumption}

Because the energies are unbounded, a tightness argument is needed
to guarantee existence. Uniqueness should be straightforward.   In what
follows, the word ``mean'' refers to expectation with respect to
$\mu_N$. We identify each site $i$ with the point $x_i =\frac{i}{N+1}$
in the unit interval, and think of $x_0=0$ and $x_{N+1}=1$ as the
locations of the baths.  For $i=1, \cdots, N$, define
\begin{align*}
\rho(x_i) &=\mbox{mean number of jumps from site $i$ to site $i+1$
per unit time}\\
&=\mbox{mean number of jumps from site $i$ to site $i-1$
per unit time;}\\
q(x_i) &=\mbox{mean energy flow from site $i$ to site $i+1$
per unit time}\\
&=\mbox{mean energy flow from site $i$ to site $i-1$
per unit time}.
\end{align*}
That is to say, the mean number of jumps out of site $i$ per unit time
is $2\rho(x_i)$; half go to the right and half to the left, and so on.
The quantities $\rho$ and $q$ are well-defined in steady state and have
very simple behavior:

\begin{lemma}\label{lemma1.1}  {\cite{ey}} With $\rho$ and $q$ defined as above, 
we have, for $x=x_i, i=1, \cdots, N$,
\begin{enumerate}
\item $\rho(x) \ = \ \rho_L +(\rho_R-\rho_L)x$;
\item $q(x) \ = \ \rho_L T_L  + (\rho_R T_R -\rho_L T_L )x$.
\end{enumerate}
\end{lemma}
As $N \to \infty$, these functions converge (trivially) to linear or,
more accurately, affine functions on $(0,1)$.  Note that $\rho'$ and
$q'$ are the steady-state tracer and energy currents, respectively.
These currents by themselves, however, do not determine steady state
profiles such as those for temperature and tracer density. The following
assumption is used to deduce such information:

\begin{assumption} {\bf (A version of LTE)} \ 
For $1 \leq i \leq N$, let $\mu_{N, i}$ denote the marginal of $\mu_N$
at the site $i$.  We assume that for every $x \in (0,1)$, $\mu_{N, [x
N]}$ converges as $N \to \infty$ to $\mu^{T, \rho(x)}$ for some $T=T(x)
>0$. A tightness condition for all the $\mu_{N, [x N]}$ is also assumed.
\end{assumption}

Let $k_i$ denote the number of tracers present at site $i$. 
Recall that $S_i$ is the stored energy.

\begin{theorem}\label{thm1} {\cite{ey}}
Let arbitrary boundary conditions $T_L, T_R, \rho_L, \rho_R$ be
given. Under Assumptions 1 and 2, 
\begin{equation}
T(x) =\lim_{N\to\infty}\mathbb{E}_{\mu_N}(S_{[xN]})\qquad\mbox{ and }\qquad
\kappa(x) =\lim_{N\to\infty}\mathbb{E}_{\mu_N}(k_{[xN]})
\end{equation}
are well-defined and are given by
\begin{equation}
T(x) = \frac{q(x)}{\rho(x)}\qquad\mbox{ and }\qquad
\kappa(x) = 2 \sqrt \pi \cdot \frac{\rho(x)}{\sqrt{T(x)}}.
\end{equation}
\end{theorem}
See {\cite{ey}} for details.

\section{Global Fluctuations and Pair Correlations}

This paper concerns the spatial energy covariances at steady state of
the random halves model described in Sect.~\ref{sec:ey-desc}.  More
precisely, let boundary conditions $T_L, T_R, \rl$ and $\rr$ be
specified, and consider an $N$-chain.  The quantities of interest are
$$
\mbox{Cov}_N(S_i,S_j) = \mathbb{E}_{\mu_N}(S_i S_j) - 
\mathbb{E}_{\mu_N}(S_i) \ \mathbb{E}_{\mu_N}(S_j),
\qquad 1\leq i,j \leq N, \ \ i \neq j.
$$
We begin by examining how these quantities scale with $N$.

\subsection{Scaling of Total-Energy Variance}
\label{sec:total-var}

For an $N$-chain with boundary conditions $T_L, T_R, \rho_L$ and $\rho_R$,
we consider the total energy variance
\begin{equation}
V(N; T_L, T_R, \rho_L, \rho_R) = \var\Big(\sum_{i=1}^N S_i\Big).
\end{equation}
Keeping the boundary conditions fixed, we treat $V=V(N)$ as a function
of $N$, and observe that $V=V_0+V_1$ where
\begin{equation}
V_0 =\sum_{i=1}^N\var(S_i)\mbox{\qquad and\qquad}
V_1 =\sum_i\sum_{j\neq i}\mbox{Cov}_N(S_i,S_j).
\end{equation}
By Theorem 1, 
\begin{equation}
\lim_{N \to \infty}\frac{1}{N} V_0(N) =\int_0^1 T(x)^2\ dx.
\end{equation}
That is, $V_0(N)\sim B_0 N$ for $N\gg 1$ with $B_0 =\int_0^1 T(x)^2\
dx$.

When $T_L=T_R$, $V_1\equiv 0$ (see Sect.~\ref{sec:ey-equilibrium}), so
that $V(N)= V_0(N) + V_1(N)\sim N$. We expect this to be true when
$\tl\neq\tr$ on physical grounds, which leads to the question: is it
true that $V_1(N)\sim B_1N$, and if so, what is the sign of $B_1$? Two
sets of numerical simulations are performed to resolve this. In one, we
compute $V_1= V-V_0$ directly as a function of $N$.  In the other, we
start from equilibrium, and hope to observe a quadratic response as a
temperature gradient is introduced; the sign of the quadratic term is
then the sign of $B_1$.  The reason we expect a quadratic (as opposed to
linear) response is symmetry: the temperature gradients $\tr-\tl=\Delta
T$ and $\tr-\tl=-\Delta T$ clearly lead to the same energy variances and
covariances.

\begin{figure}
\begin{center}
\putgraph{bb=0in 0in 2.5in 2in}{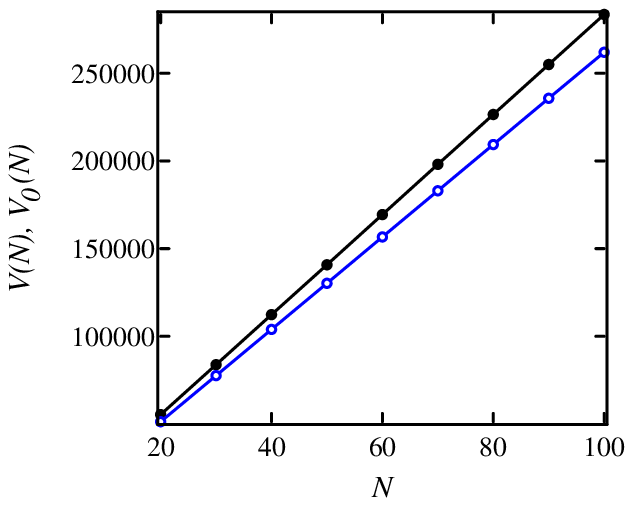}%
\end{center}
\kaption{Variance of total energy as function of system size.  The upper
curve is $V(N)$, the lower curve is $V_0(N)$.  The boundary conditions
are $T_L=10$, $T_R=100$, $\rho_L=20$, and $\rho_R=10$.}
\label{fig-total}
\end{figure}

\begin{nexp}
\label{exp1}
Fix bath temperatures $\tl,\tr$ and injection rates $\rl,\rr$ and
compute $V(N)$ and $V_0(N)$ for increasing $N$.
\end{nexp}
\noindent
The results, plotted in Fig.~\ref{fig-total}, show that
\begin{equation}
V(N)\sim BN\qquad {\rm with} \qquad B > B_0 > 0,
\end{equation}
{\em i.e.} $V_1(N) \sim B_1N$ for some $B_1>0$.

\begin{nexp}
\label{exp2}
Fix $N$, $\rl=\rr$, and a number $m > 0$.  Compute $V$ and $V_0$
for various pairs $(\tl,\tr)$ chosen so that $\frac{\tl+\tr}{2}=m$, and
investigate $V_1=V-V_0$ as a function of $\Delta T=\tr-\tl$.
\end{nexp}
\noindent
The results show that $V_1$ depends quadratically on $\Delta T$ with a
strictly positive coefficient.  Taken together, these two simulations
suggest that for $\rl=\rr$ and fixed mean temperature,
\begin{equation}
\label{eqn:total-var-scaling}
V_1(N; T_L, T_R, \rho_L, \rho_R)\propto(\tr-\tl)^2\cdot N\ ,
\end{equation}
at least when $|\tr-\tl|$ is not too large compared to the mean
temperature in the chain.  We will see later that both the linear
scaling of $V_1$ with system size and its quadratic response to
temperature gradient are consistent with the scaling of
$\cov_N(S_i,S_j)$.

Note that one advantage of probing spatial correlations via global
quantities such as $V_1$ is that it can be computed more reliably than
small, local quantities such as $\mbox{Cov}_N(S_i,S_j)$.

\subsection{Pair Covariances}
\label{sec:pair-covs}

We begin now to investigate the individual terms in the sum
$V_1=\sum_i\sum_{j\neq i}\cov_N(S_i,S_j)$.

The two most basic characteristics of $\cov_N(S_i,S_j)$ are its {\it
sign} and {\it order of magnitude}.  The fact that $V_1(N) \sim N$
suggests that many of the covariances are positive.  Indeed, all of our
numerical evidence points to $\cov_N(S_i,S_j) \geq 0$, as does the
theoretical reasoning in the sections to follow. There is reason to
remain cautious, however: Since $\cov_N(S_i,S_j)=0$ when the system is
in equilibrium (Proposition 1.1), and can be zero even when there is a
tracer flux (Proposition 5.1), one cannot conclude definitively via
numerics alone that there are no strictly negative correlations. See the
remark at the end of this subsection on why the nonnegativity of
covariances in this model may be a delicate question.

We proceed to an analysis of the order of magnitude of
$\cov_N(S_i,S_j)$, assuming in the heuristic discussion 
below that $\cov_N(S_i,S_j) \geq 0$ for all $i,j$.

We propose to decompose $V_1$ into
$$
V_1 \ = \  \sum_i V_{1,i}
\qquad {\rm where} \qquad  
V_{1,i} \ = \ \sum_{j \neq i} \cov_N(S_i,S_j)
$$
and reason as follows: 
\begin{enumerate}

\item For $i$ away from the two ends of the chain, $V_{1,i} \sim 1$
  since these terms sum to $V_1 \sim N$ and it is unlikely that they
  scale differently with $N$. (For $i$ close to 
  the boundary, $V_{1,i}$ may be smaller 
  due to the entrance of tracers  with {\it i.i.d.}\ energies.)

\item For each fixed $i$, the function $j\mapsto\cov_N(S_i,S_j)$ should
  decrease monotonically as $|i-j|$ increases: sites farther apart are
  expected to be less correlated because they ``communicate'' via longer
  and noisier ``channels.''

\end{enumerate}
These considerations imply that $\cov_N(S_i,S_{i+1})$, with $i$ away
from $0$ and $N$, are among the larger of the $\sim N^2$ terms in $V_1$.
Since $\cov_N(S_i,S_{i+1})$ is one of $N$ terms in $V_{1,i}$, (i) above
implies
$$
\cov_N(S_i,S_{i+1}) \gtrsim \frac{1}{N}\ .
$$

We point out that the order of magnitude of $\cov_N(S_i,S_{i+1})$
contains a fair amount of information about the shape of the function
$j\mapsto \cov_N(S_i,S_j)$: if $\cov_N(S_i,S_{i+1}) \sim \frac{1}{N}$,
then $\cov_N(S_i,S_j)\sim \frac{1}{N}$ for a definite fraction of $j$
(because these terms have to add up to $\sim 1$).  This points to an
extremely slow decay of $\cov_N(S_i,S_j)$ with increasing distance
between sites $i$ and $j$.  If, on the other hand, $\cov_N(S_i,S_{i+1})
\gg \frac{1}{N}$, then the function $j\mapsto \cov_N(S_i,S_j)$, $j \neq
i$, would have a very sharp peak at $j=i\pm1$.  The question is: which
scenario is the case here?

To resolve this issue, we perform the following numerical simulation:

\begin{figure}
\begin{center}
\putgraph{bb=0in 0in 3in 1.75in}{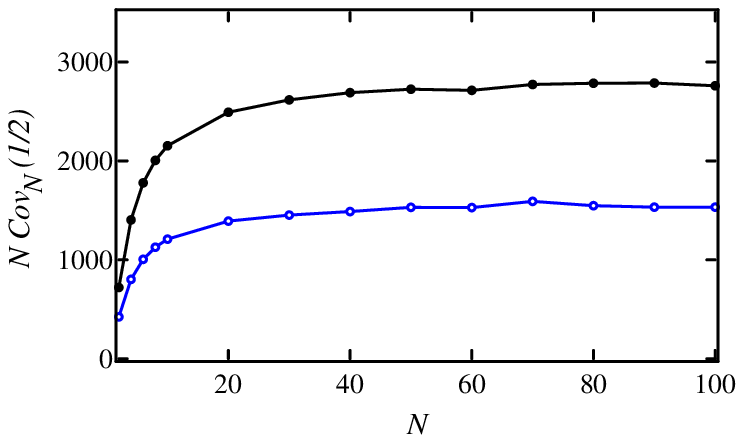}
\end{center}
\kaption{Scaling of $\cov_N$ with $N$.  This plot shows
  $N\cdot\cov_N(S_{[N/2]},S_{[N/2]+1})$ as a function of $N$.  Boundary
  conditions are $\tl=10$, $\tr=190$, and $\rl=\rr=20$ for the upper
  curve and $\rl=\rr=40$ for the lower one.}
\label{fig-middle-n}
\end{figure}

\begin{nexp}
\label{exp3}
For fixed boundary conditions, compute 
$\cov_N(S_{[N/2]},S_{[N/2]+1})$ for a range of 
$N$.
\end{nexp}
\noindent
Two sets of results are plotted in Fig.~\ref{fig-middle-n}.  They (and
other data sets not shown here) show that
\begin{equation}
\mbox{Cov}_N(S_{[N/2]},S_{[N/2]+1})\sim\frac{1}{N}.
\end{equation}
As explained above, one may infer from this that all pair
covariances $\cov_N(S_i,S_j)=O(1/N)$, a fact confirmed 
by many subsequent
simulations (see {\em e.g.} Fig.~\ref{fig-renorm}).

\paragraph{Remarks on sign of $\cov_N(S_i,S_j)$.}
As far as we know, existing analytic techniques for proving or
disproving positivity of correlations do not apply to the random halves
model.  This, in part, is due to the fact that in our model, mechanisms
conducive to both positive and negative correlations are at work, and
the actual covariance reflects a balance between these two tendencies.
For instance, a large, upward surge in the energies of incoming
tracers has the effect of raising some of the stored energies above 
their expected values. It is believed that such excursions lead to
long-wavelength fluctuations with slow relaxation times, resulting
in positive correlations~\cite{forster}.  However, upon interacting 
with one of the
tanks, a tracer that acquires a higher-than-normal fraction of the
energy is likely to move more quickly to a neighboring site and 
to interact with the tank there, possibly transferring its energy 
to the tank in its new location.  Such
a phenomenon creates negative correlations between neighboring sites.
In equilibrium, these two tendencies balance perfectly, giving zero
covariance (see Proposition {\ref{prop2.1}}).  Further numerical
evidence in support of nonnegative covariances is given in
Sect.~\ref{sec:local-grad}.


\section{Covariance Structures ``at $N = \infty$''}
\label{sec:cov-struct}

The purpose of this short section is to make explicit the objects 
to be studied in the rest of this paper and the assumptions under 
which we plan to operate. 

The discussion in the Sect.~\ref{sec:pair-covs} points to the following
functions: Given boundary conditions $\tl,\tr, \rl$ and $\rr$, and $x, y
\in (0,1)$, $x \neq y$, define ${\cal C}(x)$ and ${\cal C}_2(x,y)$ by
\begin{eqnarray*}
{\cal C}(x) & = & {\cal C}(x; \tl,\tr,\rl,\rr) \ = \ \lim_{N \to \infty} \ 
N \cdot \cov_N(S_{[xN]},S_{[xN]+1}),\\
{\cal C}_2(x,y) & = & {\cal C}_2(x,y; \tl,\tr,\rl,\rr) \ = \
\lim_{N \to \infty} \ N \cdot \cov_N(S_{[xN]},S_{[yN]}),
\end{eqnarray*}
if these limits exist.

Supposing these functions are well defined, it is natural to ask (i) how
${\cal C}(x)$ depends on local thermodynamic quantities such as $T(x)$,
$\kappa(x)$ and their gradients, and (ii) if these local quantities
completely determine ${\cal C}(x)$.  We present below an answer to the
second question; it is one of a number of possible answers.  Question
(i) is addressed in the next two sections.

Given $x \in (0,1)$ and $(\tl,\tr,\rl,\rr)$, Theorem 1 gives 
a quadruple $(T_*, \rho_*, T'_*, \rho'_*)$ where $T_*=T(x)$,
$\rho_*=\rho(x)$, $T'_*=T'(x)$ and $\rho'_*=\rho'$. 
Conversely, given $x \in (0,1)$ and a quadruple 
$(T_*, \rho_*, T'_*, \rho'_*)$, it is
easy to check that one can solve the equations in Theorem 1
backwards and find the corresponding boundary conditions.
This one-to-one correspondence between boundary and local
conditions defines a function
$\cal F$ with 
\begin{equation}
{\cal F}(x;T_*, \rho_*, T'_*, \rho'_*) = {\cal C}(x; \tl,\tr,\rl,\rr)\ .
\label{c-phi}
\end{equation}

Since we work only with finite chains, it is necessary to have a version
of ${\cal F}$ before the infinite-volume limit.  Given
$(\tl,\tr,\rl,\rr)$, $x$ and $N$, we let $i=[xN]$, and let $T_i, \rho_i,
\Delta T_i$ and $\Delta \rho_i$ be as follows: $T_i$ is the mean
temperature at site $i$, $\rho_i$ is the ``injection rate'' (see
Sect.~\ref{sec:ey-eqs} for precise definition), $\Delta
T_i=T_{i+1}-T_i$ and $\Delta \rho_i = \rho_{i+1}-\rho_i$.  Let
$$
{\cal F}_N\left(x; T_i, \rho_i, \frac{\Delta T_i}{\Delta_N}, 
\frac{\Delta \rho_i}{\Delta_N}\right) = N \cdot  \cov_N(S_i,S_{i+1}) 
$$
where $\Delta_N =\frac{1}{N+1}$. The next lemma shows that 
the function ${\cal F}_N$ is well defined:

\begin{lemma}\label{lemma3.1}
Given $N,  i,  T_i,\rho_i,\Delta T_i$ and $\Delta \rho_i$,
there is a unique set of
boundary conditions $(T_L, T_R, \rho_L, \rho_R)$ for the $N$-chain that
leads to these values at site $i$ provided $N$ is sufficiently large (or
$\Delta T_i$ and $\Delta \rho_i$ are sufficiently small).
\end{lemma}

\begin{proof}
First, $\rho_i$ and $\Delta \rho_i$ determine $\rho_L$ and $\rho_R$ by
linearity of $\rho$ (see Lemma {\ref{lemma1.1}}, {\S\ref{sec:ey-eqs}}),
provided $\Delta \rho_i$ is small enough that both of these numbers are
nonnegative.  For definiteness, assume $\Delta T_i>0$.  We will find
suitable $T_L$ and $T_R$ by varying both until correct values are
attained at site $i$: Fix the left bath temperature temporarily at $\hat
T_L \in (0,T_i)$, and vary the right bath temperature from $T_i$ to
$\infty$.  Since the temperature at site $i$ increases strictly
monotonically with the right bath temperature, there exists a unique
$\hat T_R= \hat T_R(\hat T_L)>T_i$ giving the correct value of $T_i$ at
site $i$.  Now look at the chain with bath temperatures $\hat T_L$ and
$\hat T_R$.  As $\hat T_L$ increases to $T_i$, $\Delta T_i$ decreases to
$0$, so for $\Delta T_i$ sufficiently small, there is a unique $T_L$ for
which $T_{i+1}-T_i$ is equal to $\Delta T_i$.
\end{proof}

With ${\cal F}$ and ${\cal F}_N$ as above, and for fixed boundary
conditions, the limit in the definition of ${\cal C}(x)$ is
equivalent to
\begin{equation}
\lim_{N \to \infty} {\cal F}_N
\left(x; T_{i}, \rho_{i}, \frac{\Delta T_{i}}{\Delta_N}, 
\frac{\Delta \rho_{i}}{\Delta_N}\right) = 
{\cal F}(x; T_*, \rho_*, T'_*, \rho'_*)\ .
\label{phi-phiN}
\end{equation}
Such a limit involves difficult issues beyond the scope of this paper.
For example, while the convergence of $T_i$ to $T_*$ as $N \to
\infty$ follows from LTE (Assumption 2), the convergence of
$\frac{\Delta T_{i}}{\Delta_N}$ to $T'_*$ cannot be deduced from
previous assumptions.

In addition to Assumptions 1 and 2 in Sect.~\ref{sec:ey-eqs},
we now introduce two other sets of assumptions on which the rest 
of our analysis relies. 

\begin{assumption}
The functions ${\cal C}$ and ${\cal C}_2$ are well defined and
finite-valued.
\end{assumption}

\begin{assumption}
\begin{enumerate}

\item In the context of (\ref{phi-phiN}), $\frac{\Delta T_i}{\Delta_N}
\to T'_*$ as $N \to \infty$.

\item The function ${\cal F}$ defined by (\ref{c-phi}) is differentiable.

\end{enumerate}
\end{assumption}
\noindent
Assumption 3 asserts the presence of a well-defined structure at
``$N=\infty$'' that governs the covariance relationships of the invariant
measures $\mu_N$ for large $N$.  Such a structure goes beyond the idea
of LTE to treat information of the next order, namely how $\mu_N$
deviates from local equilibrium at microscopic length scales, and how it
deviates from products of Gibbs measures (at varying local temperatures)
globally.  Assumption 4 identifies some technical issues which we take
for granted.


\section{Covariances at Microscopic Distances: $\rl=\rr$}
\label{sec:nn-equal-rate}

This section and the next concern nearest-neighbor covariances in long
chains, {\it i.e.} $\cov_N(S_i, S_{i+1})$ for large $N$.  We treat in
this section the simpler case where there is no tracer flux in the
system; the equality $\rl=\rr=\rho$ is assumed throughout.  Our aim is 
to discover how the functions $\cal C$, equivalently $\cal F$, depend
on the various quantities. 

\subsection{The two middle sites}
\label{sec:middle-sites}

In this subsection, we fix $x =\frac{1}{2}$, and study the function
$(T,\rho,T') \mapsto {\cal F}(\frac{1}{2}; T,\rho,T',0)$ where ${\cal
F}$ is as defined in Sect.~\ref{sec:cov-struct}.  Here $T, \rho$ and
$T'$ are quantities associated with the midpoint of the chain; in
particular, $T'$ is the temperature gradient at $x=\frac{1}{2}$.  We
begin by examining the dependence of ${\cal F}$ on $T'$.  Fix $T$ and
$\rho$, and consider
\begin{equation}
F(T') = F^{T,\rho}(T') = {\cal F}\big(\frac{1}{2}; T, \rho,T',0\big)\ .
\label{F}
\end{equation}
That is to say, $F(T') \approx N \cdot \cov_N(S_{[N/2]},S_{[N/2]+1})$
for large $N$ with boundary conditions $\tl= T-\frac{1}{2}T', \ 
\tr = T+\frac{1}{2}T'$ and $\rl=\rr=\rho$.
Clearly, $F(0)=0$, and $F$ is an even function due to the left-right
symmetry. This leads one to expect a quadratic
response when the system is taken out of equilibrium.
The following simulation confirms that the coefficient of the
quadratic term is indeed nonzero.

\begin{figure}
\begin{center}
\putgraph{bb=0in 0in 2.5in 2in}{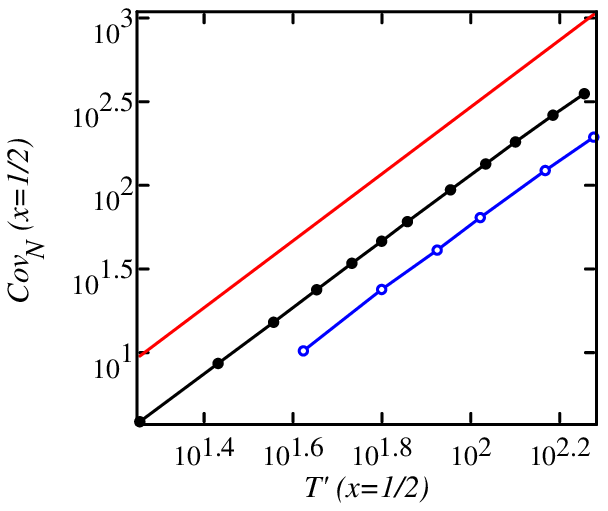}
\end{center}
\kaption{Scaling of middle-pair covariance with temperature gradient.
The plot shows $\cov_N(S_{[N/2]},S_{[N/2]+1})$ as a function of $T'$ on
a logarithmic scale.  Boundary conditions are chosen to fix $T = 60$ and
$\kappa = 2.5\sqrt\pi$ at $x=\frac{1}{2}$.  Solid discs: $N=8$; open
circles: $N=20$.  A solid line of slope 2 is shown for reference.  See
Appendix A for a discussion of simulation parameters.}
\label{fig-middle-dt}
\end{figure}

\begin{nexp}
\label{exp4}
For various pairs of $T$ and $\rho$, compute
$\cov_N(S_{[N/2]},S_{[N/2]+1})$ for a sample of $T'$.
\end{nexp}
\noindent
A set of results are plotted in Fig.~\ref{fig-middle-dt}.  The data show
that (i) $\frac{\partial^2 F}{\partial T'^2}(0) \neq 0$, and in fact,
(ii) $F(T')$ is fairly well approximated by a quadratic function over the
entire range $(-2T, 2T)$ of $T'$.  (The minimum 
temperature tends to $0$ as $|T'| \to 2T$.)
Note that if $F(T')$ is smooth, then the
only other possibility is $F(T')\sim T'^{2k}$ for some $k\in\{2,3,
\cdots\}$, which is clearly not the case here.  Other data sets (not
shown) with different values of $(T,\rho)$ confirm these conclusions.

We investigate next the dependence of $a(T,\rho) :=\frac{1}{2}\frac{d^2
F}{d T'^2}(0)$ on $T$ and $\rho$.  First, we establish a simple
result (which is expected from dimensional analysis):

\begin{lemma}\label{lemma3.4}
Fix $T$ and $\rho$, and view $\lambda >0$ as a parameter.
Define $F^{T,\rho}$ and $F^{\lambda^2T,\lambda \rho}$
as in (\ref{F}).  Then
\begin{equation}
\label{lambda-scaling}
F^{\lambda^2T,\lambda \rho}(\lambda^2 T') \ = \
\lambda^4  F^{T,\rho}(T') .
\end{equation}
\end{lemma}
\begin{proof}
Given $T,\rho$ and $T'$, let $T_L, T_R$ and $\rho_L=\rho_R$ be the
boundary conditions that give rise to these values at the two middle
sites, and consider a second system with boundary conditions $\lambda^2
T_L$, $\lambda^2 T_R$ and $\lambda \rho_L=\lambda\rho_R$ and initial
tank energies $\lambda^2$ times those of the first system.  Via a
standard coupling argument, the sample paths for these two systems are
easily matched.  Corresponding sample paths give rise to time evolutions
with identical tracer counts in both systems -- but with the second
system at energies $\lambda^2$ times that of the first and running at
speeds $\lambda$ times that of the first.  This leads to
(\ref{lambda-scaling}).
\end{proof}

Let $F^{\lambda^2T,\lambda \rho}(T') \approx a_\lambda T'^2 $.  It
follows from Lemma {\ref{lemma3.4}} that
\begin{equation}
a_\lambda\cdot (\lambda^2 T')^2 \approx 
F^{\lambda^2T,\lambda \rho}(\lambda^2 T')
\approx \lambda^4 F^{T,\rho}(T') \approx \lambda^4 a_1 T'^2.
\end{equation}
Thus $a_\lambda$ is independent of $\lambda$, and $F^{T,\rho}(T')$ can
be written as $F^\kappa(T')$ where $\kappa = 2\rho\sqrt{\pi/T}$ is the
mean tracer count with respect to $\mu^{T,\rho}$ (see
Sect.~\ref{sec:ey-equilibrium}).  The preceding argument tells us that
for any $T,\rho,\kappa$ related as above, the function $A(\kappa) :=
a(T,\rho)$ is well-defined, and
\begin{equation}
\cov_N(S_{N/2}, S_{N/2+1}) \approx \frac{1}{N} A(\kappa) T'^2\ .
\label{defA}
\end{equation}

A natural question is how $A(\kappa)$ depends on $\kappa$.  Other things
being equal, one would think intuitively that local thermal fluctuations
are smaller when more tracers are present, because stored energy is
affected less by the entrance of a tracer with very large or very small
energy.  To confirm this, and to collect data for later use, we compute
the function $A$.  Practical reasons dictate that very short chains be
used due to the large number of data points (each one of which requiring
a separate run) and the time it takes for covariances to converge in
long chains; see Appendix A for a discussion of relevant numerical
issues.  However, some understanding of the errors introduced by the use
of very short chains is necessary if these data are to be useful in
later predictions.

\begin{figure}
\begin{center}
\putgraph{bb=0in 0in 2.25in 2in}{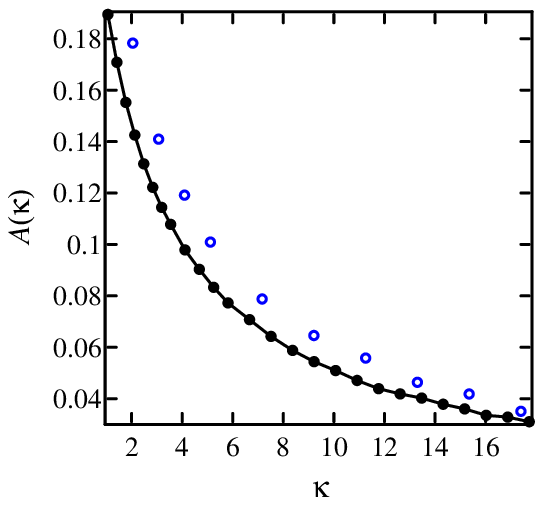}\qquad%
\putgraph{bb=0in 0in 2.4in 2in}{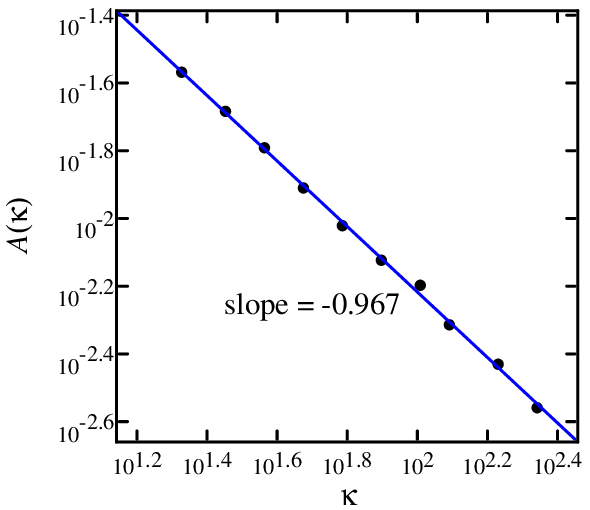}
\end{center}
\kaption{The $A$-curve computed using $8$-chains.  The boundary
conditions are $T_L=20$, $T_R=30$, and $\rho_L=\rho_R =
2.5\kappa/\sqrt\pi$, and $A(\kappa)$ is estimated using Eq.~(\ref{defA})
with $N=8$.  {\bf Left:} Linear plot.  Also shown are some points
computed using $16$-chains (open circles; $\tl=50$, $\tr=100$, and
$\rl=\rr=5\kappa/\sqrt{2\pi}$).  {\bf Right:} Log-log plot shows that
$A(\kappa)\sim 1/\kappa$ for $\kappa\gg 1$.}
\label{fig-a-curves}
\end{figure}

\begin{nexp}
\label{exp5}
Compute $A(\kappa)$ systematically for a range of $\kappa$ using chains
of 8 cells.  Compute also some values of $A(\kappa)$ using 16-chains for
comparison.
\end{nexp}
\noindent
The 8-chain results are shown in Fig.~\ref{fig-a-curves}.  Note that the
function decreases monotonically as expected, and $A(\kappa)\sim
1/\kappa$ as $\kappa\to\infty$.  

Let $A_8(\kappa)$ and $A_{16}(\kappa)$ denote the values of
$A(\kappa)$ computed using 8 cells and 16 cells respectively.  We
computed $A_{16}(\kappa)$ for some values of $\kappa$, and find these
values to be somewhat larger than the corresponding values for $A_8$.
This is consistent with Fig.~\ref{fig-middle-n}, and the true $A$-values
(defined at ``$N= \infty$'') are likely to be larger still.  We find,
however, that the {\it ratios} of the two sets of values remain fairly
constant as $\kappa$ varies.  For example,
\begin{displaymath}
1.14 < A_{16}(\kappa)/A_8(\kappa)< 1.20 \qquad {\rm for} \quad
\kappa \in \big[2\sqrt\pi, 10\sqrt\pi\big]. 
\end{displaymath}
We will use the $8$-chain $A$-curve data in our study of long-chain
covariances.  All of our long-chain simulations involve $\kappa$ in
subintervals of the range shown above, and we will only assume that
$A(\kappa)\approx\mbox{const}\cdot A_8(\kappa)$.

\subsection{Quadratic responses to local temperature gradients}
\label{sec:quad-resp}

We demonstrated in Sect.~\ref{sec:middle-sites} that at $x
=\frac{1}{2}$, the response to the local temperature gradient $T'$ is
quadratic.  We now examine the situation at $x \neq \frac{1}{2}$.
Recall that our argument for showing that the leading term has order
$\geq 2$ at $x=\frac{1}{2}$ uses the left-right symmetry of the chain, a
property not present at $x \neq \frac{1}{2}$.  Reasoning physically,
however, it is hard to imagine that the response to local temperature
gradient is sometimes quadratic, sometimes linear, or that the
coefficient of the linear term would vanish at exactly $x=\frac{1}{2}$
independent of boundary conditions.  The following simulation confirms
this thinking.

\begin{figure}
\begin{center}
\putgraph{bb=0in 0in 2.25in 2in}{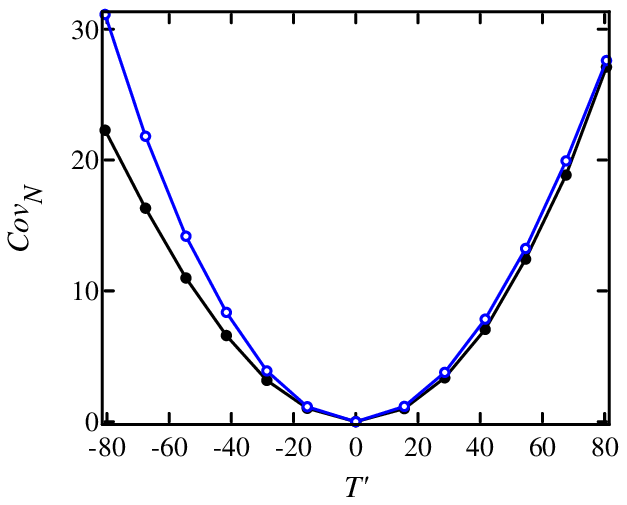}
\end{center}
\kaption{Quadratic response to $T'$ at $x \neq \frac{1}{2}$.  We fix
$x\neq 1/2$ and set $T(x)=60$, $\kappa(x)=5\sqrt\pi$.  We then compute
$\cov_N(S_{[xN]},S_{[xN]+1})$ as a function of $T'(x)$.  Solid discs:
$x\approx 0.27$; open circles: $x\approx 0.65$.  Here, we use $N=12$.
We do not take $T'$ much smaller than the values shown
for reasons discussed
in Appendix A.}
\label{fig-offcenter}
\end{figure}

\begin{nexp}
\label{exp6}
For fixed $x \in (0,1)$, $T=T(x)$ and $\rho$, compute $\cov_N(S_{[xN]},
S_{[xN]+1})$ for various values of $T'=T'(x)$.  Repeat for various $x
\neq \frac{1}{2}$.
\end{nexp}
\noindent
Fig.~\ref{fig-offcenter} shows the results for some values of $x$.
The function $T' \mapsto {\cal F}(x;T,\rho,T',0)$ for fixed $x$,
$T$, and $\rho$ is clearly quadratic to leading order and contains a
nonzero third-order term (which is not present when $x = \frac{1}{2}$).
Other sets of data, computed using different values of $T$ and $\rho$,
lead to the same conclusion.

Notice that a quadratic response to $T'$ is consistent with 
the observation in Sect.~\ref{sec:total-var} that
$\sum_{i\neq j}\cov_N(S_i,S_j) \propto(\tr-\tl)^2\cdot N$.

\subsection{Boundary effects}
\label{sec:bdry-effects}

In Sects.~4.1 and 4.2, we showed that (i) for each fixed $x$, $T$,
and $\rho$, ${\cal F}(x; T,\rho,T',0)\sim T'^2$ (at least for $|T'|$ not
too large); and (ii) at $x=\frac{1}{2}$, the coefficient in front of
$T'^2$ is a function of $\kappa(\frac{1}{2})$, that is to say,
${\cal F}(\frac{1}{2}; T,\rho,T',0) \approx A(\kappa(\frac{1}{2}))\cdot 
T'(\frac{1}{2})^2$, the function $A(\kappa)$ being {\it defined} by 
this relation at $x=\frac{1}{2}$. With $A(\cdot)$ so defined, we now ask 
if the same relation holds
at every $x \in (0,1)$, {\it i.e.}, if it is true that 
${\cal F}(x; T,\rho,T',0) \approx A(\kappa(x))\cdot T'(x)^2$.

Consider, for the moment, a different situation in which the system is
infinite in length, {\it i.e.}, $x \in (-\infty, \infty)$, and a
temperature gradient is maintained by, say, a constant external
field.  One would expect such a system to be translation-invariant,
which implies that $\cov_N(S_{[xN]}, S_{[xN]+1}) = \cov_N(S_{[yN]},
S_{[yN]+1})$ if the quantities $A(\kappa)T'^2$ at $x$ and $y$ are
identical, {\it i.e.},  we would expect the answer to the question above to 
be affirmative.

Our system has no translation invariance to speak of and, at least for
finite $N$, the randomizing effects of the baths are apparent:
nearest-neighbor covariances are significantly smaller near the
boundaries because the baths inject tracers with {\it i.i.d.} sequences
of energies into the system.  The answer to the question posed above
depends on how fast these randomizing effects dissipate.  Do they
decrease by a certain amount per lattice site, or do they decrease with
macroscopic distance?  The aim of the following simulation is to shed
light on these questions.

\begin{figure}
\begin{center}
\putgraph{bb=0in 0in 2.25in 2.25in}{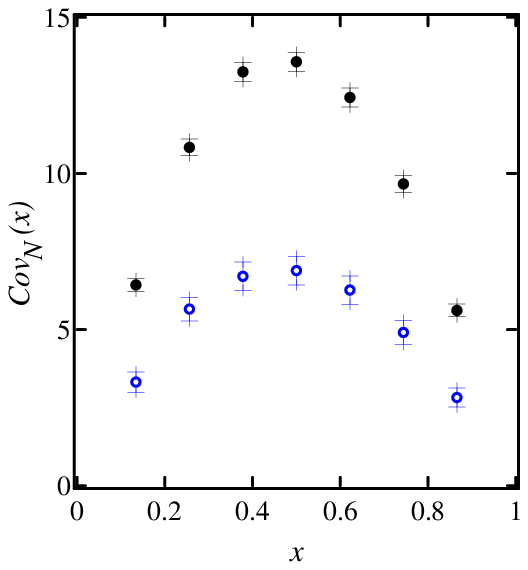}
\end{center}
\kaption{Covariances with identical local conditions at various $x$.
  For each $x$, we choose boundary conditions so that $T(x)=60$,
  $T'(x)=60$, $\rho\approx 7.75$ (closed discs) and $19.4$ (open discs).
  With these local conditions, we compute $\cov_N(S_{[xN]}, S_{[xN]+1})$
  for various $x$ symmetrically placed about $x=\frac{1}{2}$.  Note that
  each point requires a separate simulation.  Here, $N=40$.  We include
  1-standard deviation error bars; see Appendix A for factors affecting
  the numerical accuracy of computations.}
\label{fig-fixed-local-conds}
\end{figure}

\begin{nexp}
\label{exp7}
Fix a triplet $(T_*,\rho_*,T'_*)$ and compute
$\cov_N(S_{[xN]},S_{[xN]+1})$ for various $x$, using $(T_*,\rho_*,T'_*)$
as the local conditions at $x$.  That is, for each $x$, we first compute
boundary conditions which give $T(x)=T_*$, $T'(x)=T'_*$, and
$\rho(x)=\rho_*$, and then use these boundary conditions to compute the
pair covariance at $x$.  Repeat for various choices of
$(T_*,\rho_*,T'_*)$.
\end{nexp}
\noindent
The results, shown in Fig.~\ref{fig-fixed-local-conds}, show clearly
that the randomizing effects of baths dissipate in a manner more like a
diffusion process than by a definite amount per lattice site.  Note also
the asymmetry of the data points: identical local conditions at $x$ and
$1-x$ do not produce equal covariance relations.  This suggests that
from the conduction point of view, effective distance from the boundary
may not equal physical distance.

To capture the phenomenon observed, we introduce, for given boundary
conditions $\tl, \tr$ and $\rl=\rr=\rho$, a function $\varphi^{bc}(x)$
defined by the expression
\begin{equation}
{\cal C}(x) = \varphi^{bc}(x) \cdot
A(\kappa(x)) T'(x)^2\ , \qquad x \in (0,1)\ .
\label{defphi}
\end{equation}
We add the superscript ``bc'' to stress the fact that this function
depends on boundary conditions.  By definition,
$\varphi^{bc}(\frac{1}{2}) \to1$ as $T'(\frac{1}{2})\to 0$.  Since we do
not know of any other forces acting on the system, we think of
$\phibc(x)$ as a measure of the randomizing effects of the baths at
location $x$, and continue our investigation based on this thinking.

We now seek to identify the function $\phibc(x)$.  To capture the idea
of ``effective distance to boundary,'' we introduce the following {\it
time-to-boundary} functions:\footnote{Implicit here is the idea of a
dual particle system similar to that for {\it e.g.} the KMP model.  We
do not know if such a dual system can be constructed for the random
halves model.} in an $N$-chain, let $\tau^{(N)}_i$ denote the expected
time for a tracer at site $i$ to reach one of the baths.  We then have
the relation
\begin{equation}
\tau^{(N)}_i = \frac{1}{\sqrt T_i} + \frac{1}{2} \left(\tau^{(N)}_{i-1} + 
\tau^{(N)}_{i+1} \right) 
\end{equation}
where $T_i$ is the temperature profile and $\frac{1}{\sqrt T_i}$ the
expected time a tracer spends at site $i$.  Identifying the $i$th site
with $x_i=\frac{i}{N+1} \in (0,1)$ and letting $\tau = \frac{1}{N^2}
\tau^{(N)}$, we obtain, as $N \to \infty$, the differential equation
\begin{equation}
\tau''(x)= - \frac{2}{\sqrt{T(x)}}\ , \qquad \tau(0) = \tau(1) = 0 \ .
\label{tau1}
\end{equation}
We define
\begin{equation}
\phihat(x) := {\rm const} \ \tau(x)
\label{defphihat}
\end{equation}
where $\tau$ is the solution of (\ref{tau1}) and the constant is chosen
so that $\phihat (\frac{1}{2}) = \phibc(\frac{1}{2})$, and conjecture that
$\phibc$ is approximately equal to $\phihat$.
Note that $\phihat$ also depends on
boundary conditions.  We now test this conjecture numerically:

\begin{figure}
\begin{center}
\putgraph{bb=0in 0in 2.8in 2.5in}{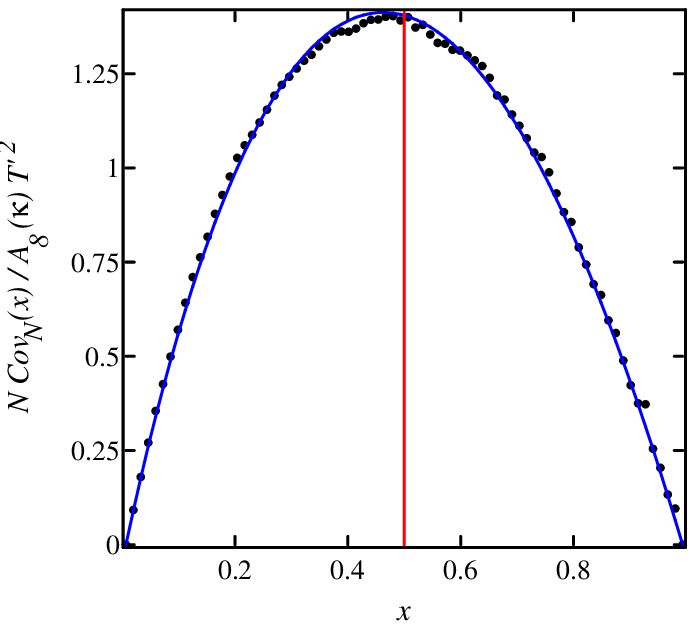}
\end{center}
\kaption{Comparison of $\phibc$ and $\phihat$.  The boundary conditions
here are $\tl=5$, $\tr=45$, $\rl=\rr=8$, and $N=60$. We estimate
$\phibc(x)$ by $N\cdot\cov_N(S_{[xN]},S_{[xN]+1}) / A_8(\kappa)T'^2$
where $A_8$ denotes the $8$-chain $A$-curve computed in
Simulation~\ref{exp5}, and plot the resulting curve against $y=
c\tau(x)$ where $c$ is chosen to minimize the (unweighted) least-squares
distance between the two curves.  A vertical line is placed at
$x=\frac{1}{2}$ for reference.  {\bf Clarifications:} (i) As noted at
the end of Sect.~\ref{sec:middle-sites}, there is a number $c_1>1$ such
that the $A\approx c_1\cdot A_8$.  Thus the data points here represent
empirical values of $c_1\phibc$.  The constant $c_1$ is absorbed into
the choice of $c$, so that the solid curve can be thought of as
$c_1\phihat$.  (ii) A least-square minimization is used instead of
setting the functions equal at $x=\frac{1}{2}$ as prescribed in the text
because the data near $x=\frac{1}{2}$ are visibly unconverged.}
\label{fig-dist-fn}
\end{figure}

\begin{nexp} 
\label{exp8}
Fix boundary conditions and $N$.  For $ x \in (0,1)$, estimate
$\varphi^{bc}(x)$ by computing $\cov_N(S_{[xN]},S_{[xN]+1})$ and
dividing the result by $\frac{1}{N}A(\kappa)T'^2$.  Compare the resulting
function to $\phihat(x)$.
\end{nexp}
\noindent
The results are displayed in Fig.~\ref{fig-dist-fn}. They show that to the
degree that we are able to estimate these functions accurately, the two graphs are in near-perfect
agreement.  These graphs are not far from a perfect parabola, but the
definitive presence of asymmetry -- in both Fig. 7 and Fig. 8 -- 
provides convincing evidence that at comparable distances,
it takes longer
to reach the bath at the lower-temperature end of the chain.  

\paragraph{Remark.}
When the system is in equilibrium, {\it i.e.}, when $\tl=\tr=T$, the
solution of (\ref{tau1}) is easily computed to be
$\tau(x)=\frac{1}{\sqrt T} x(1-x)$.  Recall that the expression $x(1-x)$
also appears in the  formulas  for
spatial covariances in the simple exclusion
{\cite{derrida1,spohn}} and KMP models {\cite{bertini2}}.  An important
difference between these models and ours is that there, the clocks signaling state changes ring at rate 1, whereas in ours, they ring at energy-dependent rates.
This property is responsible for, among other things, the slight asymmetry in
$\tau$.

\bigskip
 From the limited data available, we clearly cannot conclude the
validity of our conjecture, but details aside, the general ideas seem to
point in the right direction.


\section{Covariances at Microscopic Distances: $\rl \neq \rr$}

We return to the general case where both energy and tracer fluxes may be
present. The situation here is more complex, and we are less able to
separate the contributions of the various factors.  We will focus on the
dominant effects, and make some observations along the way.

\subsection{Responses to local gradients}
\label{sec:local-grad}

As before, we consider first $x=\frac{1}{2}$. For fixed $T$ and $\rho$, 
we consider the function
$$
F(T',\rho') = F^{T,\rho}(T',\rho') = {\cal F}(\frac{1}{2}; T, \rho,T', \rho')\ 
$$
and begin with a few observations on which terms may be absent in its
Taylor series.

\begin{proposition}\label{lemma4.2} 
$F(0,\rho') \equiv 0$ for all $\rho'$.
\end{proposition}
 
\begin{proof} Using the notation in Sect.~\ref{sec:ey}, we
claim that when
$T'=0$, the invariant measure $\mu_N$ of the chain is the product
$$
\mu^{T,\rho_1}\times\mu^{T,\rho_2}\times\cdots\times\mu^{T,\rho_N}
$$ where $\rl= \rho - \frac{1}{2}\rho'$ and $\rho_i = \rl +
\frac{i}{N+1}\rho'$ for $i=1, \cdots, N$, so that stored energies at
distinct sites are uncorrelated.  This is proved by direct verification
(along the lines of the proof of Proposition 3.7 in {\cite{ey}}).  See
Appendix B for details.
\end{proof}

Interchanging $\tl$ with $\tr$ and $\rl$ with $\rr$, we see that
$F(T',\rho')= F(-T',-\rho')$, so that there are no odd order terms in
the polynomial approximation of $F(T',\rho')$.  Proposition {\ref{lemma4.2}}
says there are no terms that are purely powers of $\rho'$, and we have
shown in Sect.~\ref{sec:middle-sites} that among the terms that are pure
powers of $T'$, the leading one is $T'^2$.

Perturbing from equilibrium, we have shown that $\frac{\partial
F}{\partial T'}(0,0) =\frac{\partial F}{\partial \rho'}(0,0)=0$.
Consider the Hessian $D^2F(0,0)$.  Since $\frac{\partial^2F}{\partial
\rho'^2}=0$ (Proposition {\ref{lemma4.2}}), it follows that for
$|T'|,|\rho'|\ll 1$,
\begin{equation}
F(T',\rho')\approx  aT'^2 + bT'\rho', \quad  {\rm for \ some} \quad
a=a(T,\rho)\mbox{ and }b=b(T,\rho).
\end{equation}

Recall our brief discussion of the sign of $\cov_N(\cdot, \cdot)$ in
Sect.~\ref{sec:pair-covs}.  Our next lemma says that the degeneracy of
the Hessian $D^2F(0,0)$ is equivalent to all covariances at $x=\frac{1}{2}$
having the same sign, namely the sign of $a(T,\rho)$.

\begin{lemma} $F \geq 0$ in a neighborhood of $(0,0)$
if and only if $a \geq 0$ and $b=0$.
\end{lemma}
\begin{proof}
As a quadratic form, the Hessian $D^2F(0,0)$ has maxima and minima in
perpendicular directions.  We know already that it is degenerate along
the $\rho'$-axis.  Thus either the minimum is negative, that is
$F(T',\rho')<0$ for some $(T',\rho')$, or the Hessian is positive
semi-definite, {\em i.e.} $a \geq 0$ and $b=0$.
\end{proof}

\begin{figure}
\begin{center}
\setlength{\unitlength}{1in}
\begin{picture}(5,2.5)(0,0)
\put(-0.2,0){\putgraph{bb=0in 0in 1.8in 1.8in}{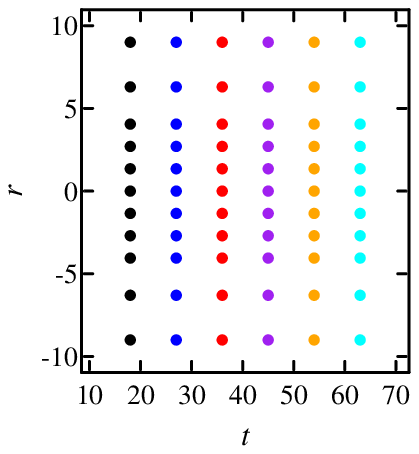}}
\put(2.2,1.1){\small\begin{tabular}{|c|c|c|}
\multicolumn{3}{c}{ }\\
\hline
Term     & Coeff. ($\rho=12.5$) & Coeff. ($\rho=20$)\\\hline\hline
$ t^2 $& 0.094 & 0.067 \\
$ tr $& $1.3\times 10^{-3}$ & $5.8\times 10^{-4}$ \\
$ t^4 $& $-1.8\times 10^{-7}$ & $-1.3\times 10^{-7}$ \\
$ t^3r $& $-3.8\times 10^{-6}$ & $-1.8\times 10^{-6}$ \\
$ t^2r^2 $& $1.1\times 10^{-4}$ & $3.0\times 10^{-5}$ \\
$ tr^3 $& $-4.5\times 10^{-5}$ & $-7.6\times 10^{-6}$ \\
\hline
\end{tabular}}
\end{picture}
\end{center}
\kaption{Local polynomial expansion of $F$.  We set $T = 100$,
$\rho\in\{12.5,20\}$, $T'=t$, and $\rho'=r$ at $x=\frac{1}{2}$, then
compute $F(t,r)\approx N\cdot\cov_N(\frac{1}{2})$ for $N=8$ using the
$(t,r)$ values from the grid ({\bf left panel}).  We then find the best
polynomial (in the sense of least-squares) of the form $at^2 + btr +
ct^4 + dt^3r + et^2r^2 + ftr^3$ which fits the data ({\bf right panel}).
For both values of $\rho$, the absolute standard error for the coefficient of the
$tr$ term is $\sim 2\times 10^{-3}$.}
\label{fig-interp2d}
\end{figure}

To determine the leading-order terms of the Taylor expansion of $F$ near
$(0,0)$, in particular to see if $b\equiv 0$ (equivalently $F\geq 0$),
we perform the following simulation:
\begin{nexp}
\label{exp9}
Fix $T$ and $\rho$, and set $T'=t$, $\rho'=r$.  Compute $F(t,r)$ for
a grid of $(t,r)$ values and use the resulting data to find the best
fitting 4th degree polynomial of the form
\begin{equation}
P(t,r) \ = \ at^2 + btr + ct^4 + dt^3r + et^2r^2 + ftr^3\ .
\end{equation}
\end{nexp}
\noindent
Two sets of results is shown in Fig.~\ref{fig-interp2d}.  We find
that in these simulations (and in others not shown here), the
leading-order term in the Taylor polynomial of $F(T',\rho')$ is $aT'^2$
with $a>0$.  In all cases, $|b|\ll 1$, which is consistent with all
observed covariances being $\geq 0$.  As expected, we cannot conclude
definitively that $b\equiv 0$.

 From here on we take as a working assumption that the quadratic form
associated with $D^2F(0,0)$ above is degenerate, {\it i.e.}, at
$x=\frac{1}{2}$, the only second order term is $A(\kappa) T'^2$ as in
the $\rl=\rr$ case. For similar reasons as before, namely that the
qualitative picture elsewhere in the chain should not differ from that
at $x=\frac{1}{2}$, we assume further the absence of $T'\rho'$-terms in
the Taylor expansion of $F$ at all $x \in (0,1)$.  This completes our
discussion of how covariance depends on local quantities when the system
is not far from equilibrium.

\subsection{Far-from-equilibrium corrections}

We have shown that at $x=\frac{1}{2}$, $F^{T,\rho}(T',\rho') \approx
A(\kappa) T'^2$ is a good approximation of the response to the local
temperature and injection gradients -- {\em provided $|T'|$ and
$|\rho'|$ are sufficiently small}.  In the $\rho'=0$ case, we noted in
Sect.~\ref{sec:middle-sites} that the function $T' \mapsto
F^{T,\rho}(T')$ is close but not equal to a perfect quadratic for large
$T'$.  We now investigate the effect of larger $|\rho'|$ on
$F^{T,\rho}(T',\rho')$.
\begin{figure}
\begin{center}
\putgraph{bb=0in 0in 3in 1.75in}{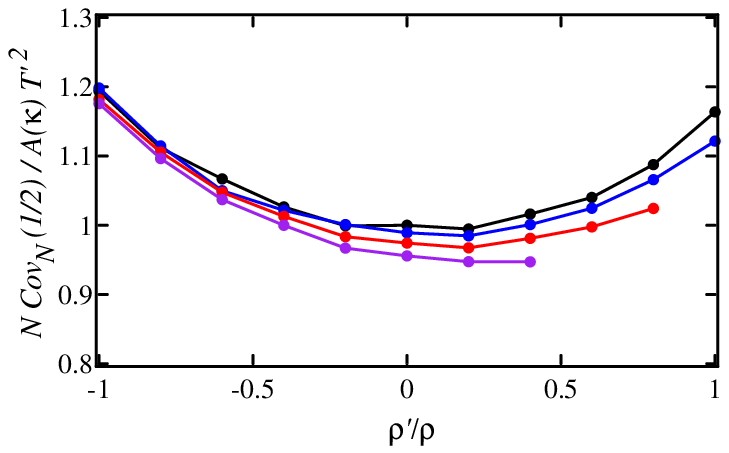}
\end{center}
\kaption{Responses to local gradients for various $T'$ and $\rho'$ at
$x=\frac{1}{2}$.  We fix $T$, $\rho$, and $T'$ at $x=\frac{1}{2}$, then
plot $N\cdot\cov_N(\frac{1}{2}) / A(\kappa) T'^2$ as a function of
$\rho'/\rho$.  Each curve corresponds to one value of $T'$.  The
parameters are $N=12$, $T(\frac{1}{2}) = 50$, 
$\rho(\frac{1}{2})\approx 9.68$, and $T'(\frac{1}{2})=90$, 70, 50, and
30 (top to bottom).  Some of the curves are discontinued
because when $T'(\frac{1}{2})$ and $\rho'(\frac{1}{2})$ are large, 
$T(x)$, which decreases monotonically with $x$, may reach 
$0$ before $x$ does.  This is easily seen from Theorem {\ref{thm1}}.}
\label{fig-far-from-eq}
\end{figure}

\begin{nexp}
\label{exp10}
For fixed values of $(T,\rho)$, compute $F^{T,\rho}(T',\rho')$ at
$x=\frac{1}{2}$ as a function of $T'$ and $\rho'$, and study the
deviations from the function $A(\kappa) T'^2$.
\end{nexp}
\noindent
A set of results is shown in Fig.~\ref{fig-far-from-eq}.  The results
for other choices of $(T,\rho)$ (not shown) are similar.

Observe that (i) the dominant factor is $A(\kappa)T'^2$, but a moderate
correction (roughly 20\% for $|\rho'/\rho| \leq 1$) to the leading
coefficient is sometimes needed; and (ii) a larger $|\rho'|$ tends to
increase covariances.  Along the lines of the conjectural thinking that
positive covariances are caused by waves of abnormally high (or low)
energy tracers with long wavelengths, the presence of a tracer flux
appears to amplify the propagation of these waves.

\subsection{Nearest-neighbor covariances in long chains}
\label{sec:long-dist-covs}

We have seen that the presence of a tracer flux complicates the behavior
of covariances at $x=\frac{1}{2}$ when the system is far from
equlibrium.  We do not know how it affects the propagation of boundary
effects, or how to separate these contributions (or if they can be
separated at all).  We will demonstrate in this subsection, however,
that the {\it main ingredients} in $\cov_N(S_i,S_{i+1})$ are those
already identified in Sect.~\ref{sec:nn-equal-rate}.

More precisely, let $\phibc$ and $\phihat$ be as defined in
Sect. \ref{sec:bdry-effects}.  That is to say, $\phibc$ is as defined in
(\ref{defphi}), $\tau$ the solution of (\ref{tau1}), and $\phihat(x) =
c\tau(x)$ where $c$ is chosen to ensure $c\tau (\frac{1}{2}) =
\phibc(\frac{1}{2})$.  We test
the validity of
\begin{equation}
\cov_N(S_{[xN]}, S_{[xN]+1}) \approx \frac{1}{N} \cdot \phihat(x)
\cdot  A(\kappa (x)) \cdot T'(x)^2
\label{nn}
\end{equation}
as an approximate relation.

\begin{figure}[!t]
\begin{center}
\subfigure[$\tl=10,\tr=50,\rl=15,\rr=9; N=70,90,120$]{\putgraph{bb=0in 0in 2in 2in}{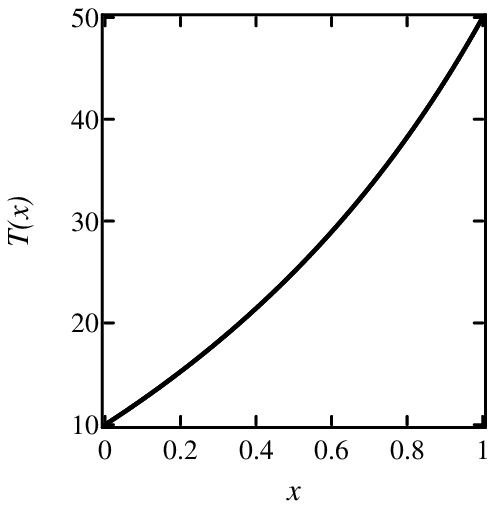}\qquad%
\putgraph{bb=0in 0in 3.25in 2.5in}{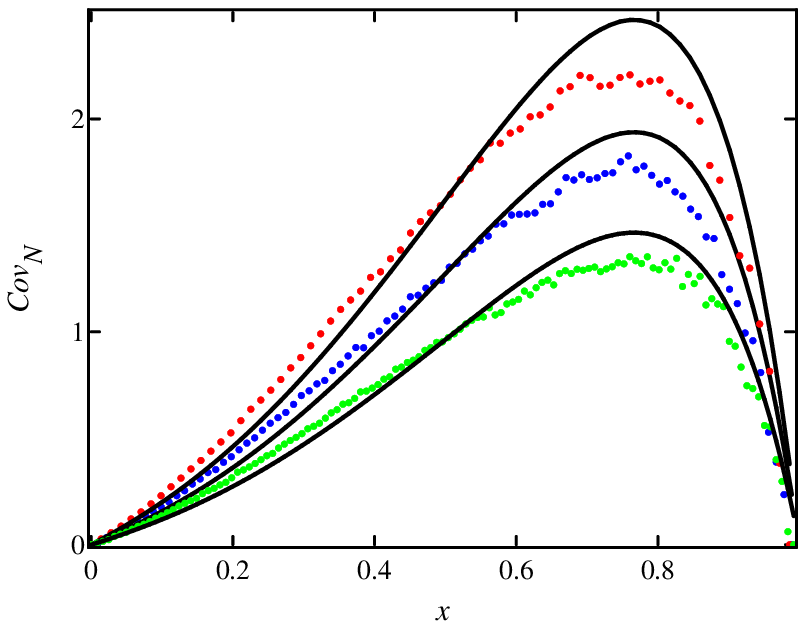}}\\
\vspace{12pt}
\subfigure[$\tl=5,\tr=45,\rl=8,\rr=15; N=50,70$]{\putgraph{bb=0in 0in 2in 2in}{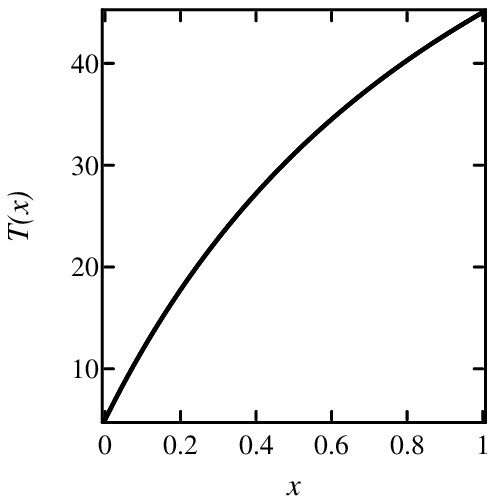}\qquad%
\putgraph{bb=0in 0in 3.25in 2.5in}{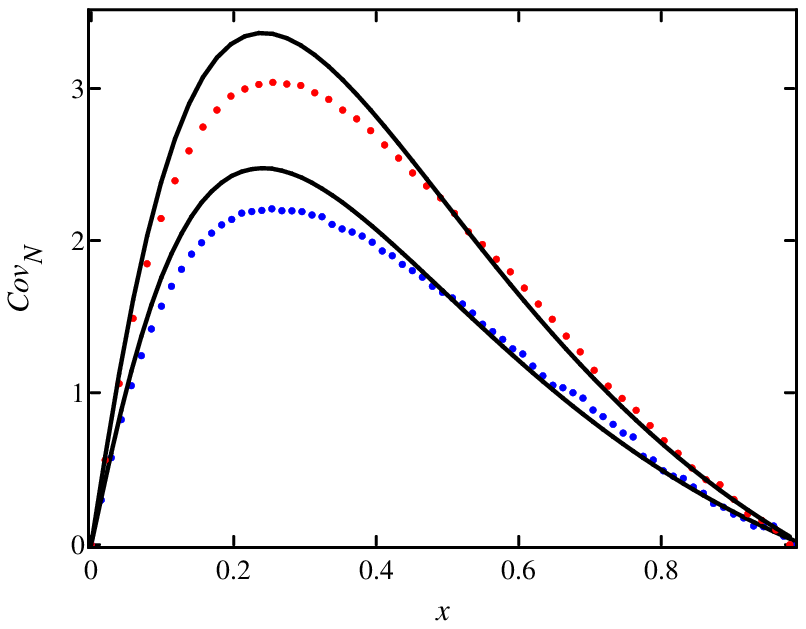}}
\vspace{12pt}
\end{center}
\kaption{Nearest-neighbor covariances.  {\bf Left column:} Temperature
profiles $T(x)$. {\bf Right column:} Empirically computed values of
$\cov_N(S_i, S_{i+1})$, plotted against predicted covariances using
Eq.~(\ref{nn}).  The predicted covariance curves are scaled to coincide
with the empirical curves at $x=\frac{1}{2}$.}
\label{fig-long-chains}
\end{figure}

\begin{nexp}
\label{exp11}
For fixed boundary conditions with $\rl\neq\rr$, and $N$ taken as large
as possible, compare empirically computed values of
$\cov_N(S_i,S_{i+1})$ to their predicted values given by (\ref{nn}).
\end{nexp}
\noindent
The results for two sets of boundary conditions are plotted in
Fig.~\ref{fig-long-chains}.  Notice that only the shapes of ${\cal
C}(x)$ are being predicted because of unknown normalization factors.

Fig.~\ref{fig-long-chains} confirms many features of our
predictions. First, the shapes of ${\cal C}(x)$ confirm that the
randomizing effects of the heat baths are diffusive in nature; these
effects result in an effective-distance factor whose graph takes the
form of a distorted parabola.  Second, the locations of the peaks of
${\cal C}(x)$ show the influence of $T'$; see the left column of Fig.11.\footnote{In Fig. 11(a), both 
$T'^2$ and $A$ increase from left
to right; in (b) $T'^2$ increases from  $4\times 10^2$ to $6\times 10^3$
while $A$ decreases from $0.07$ to $0.04$.}
Finally, comparison of the data for
different $N$ in each set provide another confirmation of the $O(1/N)$
scaling.

Fig.~\ref{fig-long-chains} also shows the need for corrective
factors on the order of $10-15\%$ away from the boundaries.  
While we do not know the precise
nature of these corrections, we point out that in the presence of a
tracer flux, the functions $\phibc$ (as computed from empirical
data) are a little more asymmetric than in the case $\rl=\rr$
(shown in Fig.~\ref{fig-dist-fn}).

\section{Long-range Covariances}
\label{sec:renorm}

\subsection{Renormalizability and the function ${\cal C}_2(x,y)$}

In Sect.~\ref{sec:cov-struct}, we introduced the idea of a
pair-covariance function ${\cal C}_2(x,y)$.  The existence of the limit
in the definition of ${\cal C}_2(x,y)$ implies the following: for all
sufficiently large $N, N'$,
\begin{equation}
\label{eqn:limit equiv}
N\cdot\cov_N(S_{[xN]},S_{[yN]}) \approx
N'\cdot\cov_{N'}(S_{[xN']},S_{[yN']})\ .
\end{equation}
This can be seen as a statement about the {\it renormalizability} of
pair covariances: Consider an $N$-chain with $N=rN_0$ for some integers
$N_0$ and $r$, and subdivide the chain into $N_0$ groups of $r$
consecutive sites. For convenience, we take $r$ to be odd. Let $i^{(r)}
= (i-1)N_0 + \frac{1}{2}(r+1)$, {\it i.e.}, $i^{(r)}$ is the index of
the middle of the $r$ sites in the $i$th group.  For any $i,j$ with $1
\leq i,j \leq N_0$ and $i \neq j$, we compare the covariance at sites
$i$ and $j$ in an $N_0$-chain to that at sites $i^{(r)}$ and $j^{(r)}$
in an $N$-chain with the same boundary conditions. Eq.~(\ref{eqn:limit
equiv}) tells us that $r\cdot\cov_{rN_0}(S_{i^{(r)}},S_{j^{(r)}})
\approx \cov_{N_0}(S_i,S_j)$ if $N_0$ is sufficiently large, and that as
$r\to\infty$, $r\cdot\cov_{rN_0}(S_{i^{(r)}},S_{j^{(r)}})$ converges to
a constant.

We test this renormalizability to confirm that ${\cal C}_2(x,y)$ is well
defined.

\begin{figure}[t]
\vspace{15pt}
\begin{center}
\putgraph{bb=0in 0in 2.5in 2in}{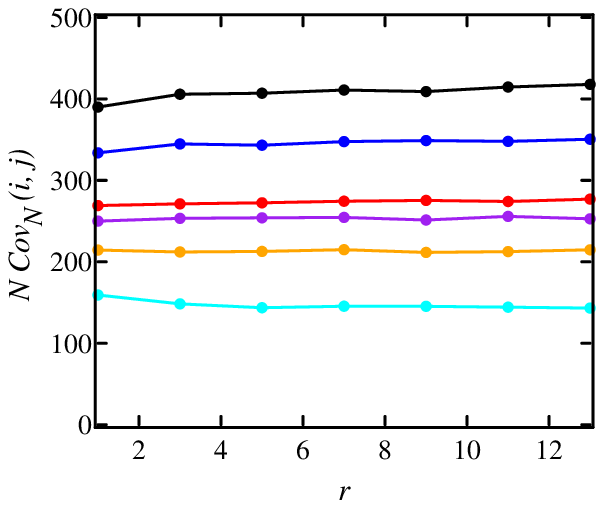}
\end{center}
\kaption{Renormalizability of the covariance function.  The plot shows
  $N\cdot\cov_{N}(S_{i^{(r)}},S_{j^{(r)}})$, where $N=rN_0$, $N_0 = 8$,
  and $r = 1,3,5,\cdots,13$.  From top to bottom, the curves correspond
  to the $(i,j)$ pairs $(3,4)$, $(3,5)$, $(2,3)$, $(2,4)$, $(2,5)$, and
  $(2,6)$.  Boundary conditions are $\tl=10,\tr=100,\rl=20,\rr=10$.}
\label{fig-renorm}
\end{figure}

\begin{nexp}
\label{exp12}
Fix $\tl$, $\tr$, $\rl$, $\rr$, and $N_0\in {\mathbb Z}^+$.  For
various pairs $(i,j)$ in the $N_0$-chain, compute
$N\cdot\cov_{N}(S_{i^{(r)}},S_{j^{(r)}})$ for $N=rN_0$,
$r=3,5,7,\cdots$.
\end{nexp}
\noindent
A subset of the data is shown in Fig.~\ref{fig-renorm}.  Because the 
value of $N_0$ used is relatively small ($N_0=8$),
one can expect the plotted values $rN_0 \cdot\cov_{N}(S_{i^{(r)}},S_{j^{(r)}})$
to converge or stabilize to constants only as $r$ increases.  
The results show that they, in fact, stabilize fairly quickly.

\begin{figure}[t]
\begin{center}
{\scriptsize\begin{tabular}{ccc}
\putgraph{bb=-.1in 0in 2in 1.8in}{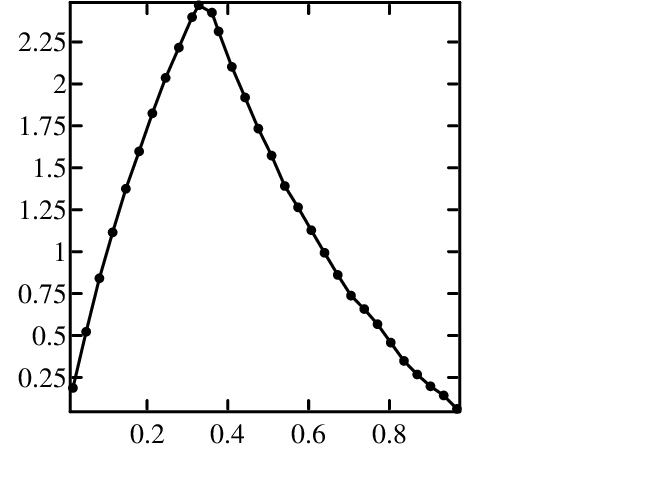}&
\putgraph{bb=-.1in 0in 2in 1.8in}{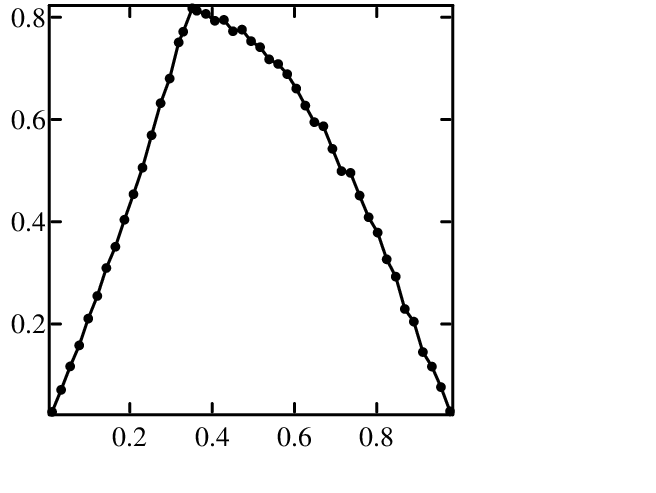}&
\putgraph{bb=-.1in 0in 2in 1.8in}{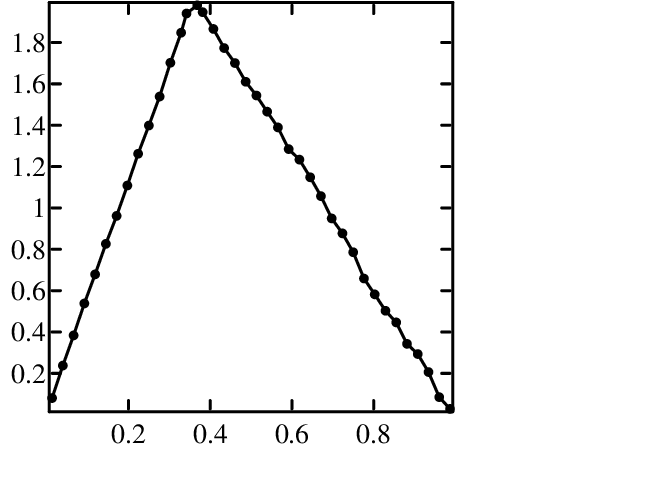}\\
(a) $\tl=5,\tr=45,\rl=8,\rr=15$;&
(b) $\tl=10,\tr=50,\rl=15,\rr=9$;&
(c) $\tl=5,\tr=45,\rl=8,\rr=8$;\\
\hspace{-90pt}$N=60$ &\hspace{-90pt}$N=90$ &\hspace{-90pt}$N=75$\\
\end{tabular}}
\vspace{6pt}
\end{center}
\kaption{Covariances at macroscopic distances.  These plots show
$\cov_N(S_{[xN]},S_{[yN]})$ as a function of $y$, with $x$ fixed at
$\approx 1/3$.}
\label{fig-long-dist-covs}
\end{figure}

That ${\cal C}_2(x,y)$ is well defined implies that the function $(x,y)
\mapsto \cov_N(S_{[xN]}, S_{[yN]})$, when normalized, settles down to a
fixed shape for large $N$. We now investigate the shapes of these
functions. While carrying out Simulations~\ref{exp8} and {\ref{exp12}},
we also collected data for $\cov_N(S_i,S_j)$ for various pairs of $i,j$.
Graphs of $j \mapsto N \cdot \cov_N(S_i, S_j)$ with $i = [\frac{1}{3}N]$
are shown in Fig.~\ref{fig-long-dist-covs} for three sets of boundary
conditions.  As predicted in Sect.~\ref{sec:pair-covs}, these functions
are bounded, and they decrease monotonically to $0$ as $|i-j|$
increases.  This decay rate is roughly linear in macroscopic distance,
and extraordinarily slow per lattice site.  For example, if $N=10^6$ and
$i=[\frac{1}{3}N]$, then $\cov_N(S_i, S_{i+1}) \approx 2 \cdot
\cov_N(S_i, S_{2i})$.  We note that these findings are consistent with
fluctuating hydrodynamics; see {\it e.g.}, \cite{spohn}.

\subsection{An approximate formula}
\label{sec:approx-long-dist}

A natural generalization of the nearest-neighbor covariance formula
(\ref{nn}) to pair covariances separated by macroscopic distances is 
\begin{equation}
\cov_N(S_{[xN]}, S_{[yN]}) \approx \frac{1}{N} \cdot \varphi^{bc}_2(x,y)
\cdot  \bar A(x,y) \cdot \bar T'(x,y)^2\ .
\label{C2}
\end{equation}
We think of this as an approximate formula that holds for
$x,y \in (0,1)$ with  $0 < |x-y|\ll 1$: $\varphi^{bc}_2(x,y)$ is a notion of effective distance of {\it the pair} $x,y$ to the boundary, and $\bar A$
and $\bar T'$ are generalizations of corresponding quantities in 
(\ref{nn}).  As a rough approximation, one may take
\begin{equation}
\bar A(x,y) = \frac{1}{2}\big(A(\kappa(x)) +A(\kappa(y))\big), \qquad \bar
T'(x,y) = \frac{T(y)-T(x)}{y-x}\ .
\label{avg}
\end{equation}

\begin{figure}
\begin{center}
\putgraph{bb=0in 0in 2.5in 2.25in}{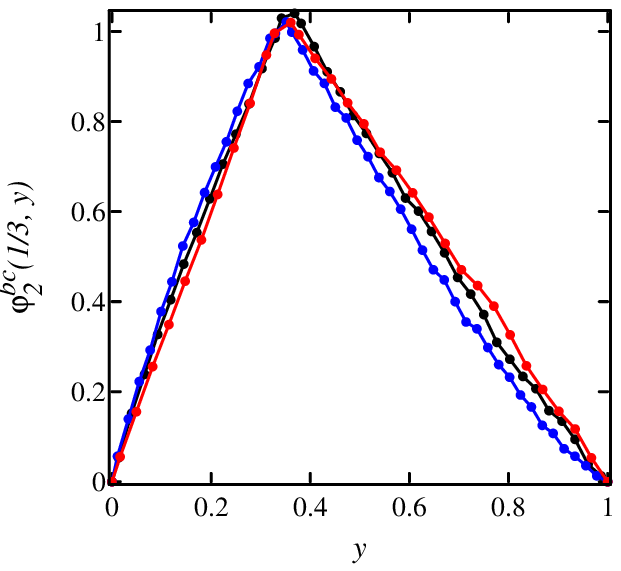}
\end{center}
\kaption{A slice of the 2-point distance function $\varphi^{bc}_2$.
  Here, we use Eqs.~(\ref{C2}) and (\ref{avg}) to compute the graphs of
  $y\mapsto\varphi^{bc}_2\big(x,y\big)$, $x \approx\frac{1}{3}$, from the data
  in Fig.~\ref{fig-long-dist-covs}.  The results are normalized to $1$
  at $x$ and overlaid.}
\label{fig-2pt-dist-fn}
\end{figure}

To obtain information on $\varphi^{bc}_2(x,y)$, we estimate it
using (\ref{C2}), taking $\bar A$ and $\bar T'$ as above.  Slices of the
graphs of these functions of 2 variables, with $x$ fixed and $y$
varying, are shown in Fig.~\ref{fig-2pt-dist-fn}.

Piecing together these slices, we deduce the following
geometric facts about our 2-point distance function
$\varphi^{bc}_2$ defined for $(x,y) \in (0,1) \times (0,1)$:
it is continuous and piecewise smooth, with a ``ridge''
along the line $\{x=y\}$.  Along that line, it is equal to
$\varphi^{bc}(x)=\varphi^{bc}(y)$, the value of the 1-point
distance function introduced in
Sect.~\ref{sec:bdry-effects}.  For fixed $x$, the function
$y \mapsto \varphi^{bc}_2(x,y)$ is roughly piecewise linear,
peaking at $y=x$.

Observe that if this function were {\em exactly} piecewise linear, then
by exchanging $x$ and $y$, we would arrive at the relation
$$
\phibc(x) \cdot \frac{1-y}{1-x} \ = \ \phibc(y) \cdot \frac{x}{y}
\qquad {\rm for} \quad x<y\ .
$$ This relation implies that $\phibc(x) = c x(1-x)$ for some constant
$c$.  Since $\phibc$ is roughly parabolic (see
Sects.~\ref{sec:bdry-effects} and {\ref{sec:long-dist-covs}}), the
approximate piecewise linearity of $\varphi^{bc}_2$ is consistent with
our results from previous sections. (Exact formulas for the simple
exclusion {\cite{derrida1,spohn}} and KMP models {\cite{bertini2}} also
contain piecewise linear functions.)  On the other hand, we know
from Sections 4 and 5 that for the random halves model, $\phibc(x) \neq
c x(1-x)$.  Thus $\varphi^{bc}_2$ cannot be exactly piecewise linear
either.

\bigskip
In summary, our results for long-range covariances are rough and are
obtained by extrapolating from what we know about covariances at
microscopic distances, together with a numerical determination of the
distance function $\varphi^{bc}_2$.  An approximate formula is
$$
\cov_N(S_{[xN]}, S_{[yN]}) \ \approx \  \begin{cases} 
\ \frac{1}{N} \cdot \left(\frac{1-y}{1-x}\right) \cdot \phihat(x) 
\cdot \bar A(x,y)
\left(\frac{T(y)-T(x)}{y-x}\right)^2
\ \ \ \ {\rm for} \ \ \ \ 0 \leq x \leq y \leq 1\ , \\
\ \frac{1}{N} \cdot \left(\frac{y}{x}\right) \cdot \phihat(x) \cdot 
\bar A(x,y)  \left( \frac{T(y)-T(x)}{y-x} \right)^2
\ \ \ \ \quad {\rm for} \ \ \ \  0 \leq y \leq x \leq 1 \ ,
\end{cases}
$$ where $\phihat$ is our approximate one-point distance function and
$\bar A$ is as in (\ref{avg}).  Both this formula and
Fig.~\ref{fig-long-dist-covs} show clearly the following:
\begin{enumerate}

\item Covariance decays essentially linearly with macroscopic distance,
{\it i.e.},
$$
{\cal C}(x) - {\cal C}_2(x,y) \ \sim \ |x-y|\ .
$$

\item For fixed $x$, the function $y \mapsto {\cal C}_2(x,y)$ is
continuous but not differentiable at $y=x$; it has opposite concavity on
the two sides of $x$ if the temperature profile $T(y)$ is nonlinear.

\end{enumerate}
That the curvature changes sign at $x$ is clearly visible in
Fig.~\ref{fig-long-dist-covs}; it is also evident from the second
derivative of $\big(\frac{T(y)-T(x)}{y-x}\big)^2$ (see
Fig.~\ref{fig-long-chains}, left column, for the temperature profiles
$T(y)$).

\section*{Conclusions and Remarks}
\addcontentsline{toc}{section}{\numberline {}Conclusions and Remarks}

Via a series of theoretical arguments and numerical simulations, we have
developed a coherent picture for the spatial covariances at steady state
of the 1-D random halves model.  We have established firmly that
stored-energy covariances have order of magnitude $\frac{1}{N}$ away
from the boundaries of $N$-chains.  This in itself points to the
presence of long-range covariances which decay very slowly.  Subsequent
analysis shows that covariances decay {\it linearly} with macroscopic
distance.  For sites separated by microscopic distances, we have a
simple formula that encapsulates the main ingredients on which energy
covariances depend, including (i) a quadratic response to local
temperature gradient, (ii) diffusive nature of the randomizing effect of
the heat baths, and (iii) stabilizing effects of large numbers of
tracers.

Since the random halves models are stochastic idealizations of certain
mechanical models, we hope our results will also shed light on these and
similar Hamiltonian systems.  There are, however, important differences,
such as mixing and memory issues (see {\em e.g.} {\cite{ey,eckmann}}).
The extent to which the picture established here carries over to the
Hamiltonian setting remains to be seen.

Finally, it is well known that nonequilibrium phenomena are quite
different in higher dimensions.  Our detailed study here provides a
baseline for explorations in 2-D and 3-D, a project currently being
carried out by the authors.  The results, which are indeed different as
predicted by fluctuating hydrodynamics, will be reported in a
forthcoming paper.


\vskip .2in
\noindent {\it Acknowledgements.} 

We are grateful to Jonathan Goodman, Leo Kadanoff, Oscar Lanford, Joel
Lebowitz, Charles Newman, Luc Rey-Bellet and David Ruelle for helpful
conversations.


\section*{Appendices}
\addcontentsline{toc}{section}{\numberline {}Appendices}
\subsection*{A\ \ \ Remarks on simulations}

The simulations carried out in this paper implement directly
the dynamics described in Sect.~\ref{sec:ey-desc};
expectation values with respect to the invariant measures are
computed via time averages over long trajectories.  The
numerical issues are quite similar to those of Markov chain
Monte Carlo computations.

We calculate empirical error bars using a standard ``batch
means'' estimator.  These error bars measure only
statistical errors that arise from the fact that
expectation values are estimated by time-averaging over
finite intervals of time, {\em i.e.} finite-length
trajectories.  The error bars do not account for {\em
finite-size effects}, {\em i.e.} bias due to the fact that
we can only simulate $N$-chains with finite $N$.  To improve
clarity and readability, we have suppressed the error bars in
most of the figures, displaying only those that are directly
relevant to the issues being discussed.

A variety of variance reduction techniques have been
invented to speed up the convergence of Monte Carlo
calculations, ranging from multigrid Monte Carlo
{\cite{sokal}} to large deviations-based importance sampling
{\cite{bucklew}}.  Most of these techniques require
additional information, such as an explicit expression for
the invariant measure or a large deviations functional.
Such methods do not work in our setting.  There is on-going
work on a class of algorithms which do not require detailed
knowledge of the invariant measure {\cite{goodman}}.
However, it is not known whether such techniques can be
applied to the random halves model.

The quantities of interest in most of our simulations are the
$\cov_N(S_i, S_j)$.  In general, these are relatively small numbers that
are differences of two much larger numbers and can be rather costly to
compute.  We estimate the cost of computing $\cov_N$ as follows: for
temperatures which are $\sim T$, both ${\mathbb E}(S_i S_j)$ and
${\mathbb E}(S_i) {\mathbb E}(S_j)$ are $\sim T^2$, while $\cov_N(S_i,
S_j)\sim A(\bar\kappa) T'^2/N$, where $\bar\kappa$ is the typical tracer
density per site.  (For the present discussion, we focus on $i,j$ away
from the boundaries so that $\phibc(x)\sim 1$.)  In order to compute
$\cov_N(S_i,S_j)$ with a relative error of $\eps$, we need ${\mathbb
E}(S_i S_j)$ with a relative error of $\eps A(\kappa) (T'/T)^2 / N$; the
same is true for ${\mathbb E}(S_i) {\mathbb E}(S_j)$.  Since the
statistical error in the time average $\frac{1}{\tau}\int_0^\tau
S_i(\tau') S_j(\tau')\ d\tau'$ is $\sim\sqrt{\alpha/\tau}$ for some
$\alpha$, this means we need to integrate the system for a time $\tau$
which is proportional to $N^2/\eps^2 A(\bar\kappa)^2\cdot(T/T')^4$.
Now, the computational cost of simulating the system up to time $\tau$,
as measured by the total number of ``events,'' is proportional to the
number of tracers $\bar\kappa N$ in the system and the mean rate of
activity of each tracer (which is $\sim\sqrt{T}$).  Thus, we have
\begin{equation}
\label{eqn:runtime}
\mbox{computational cost}\sim \alpha \cdot \frac{N^3\bar\kappa}{\eps^2
  A(\bar\kappa)^2} \left(\frac{T}{T'}\right)^4\sqrt{T}\ .
\end{equation}
For our simulations involving long chains, we have found $\alpha$ to be 
typically $\lesssim 10$. As an example, for $\eps = 5\%$, $N=60$, 
$\tl=5$, $\tr=45$, $\rl=\rr=8$,
we have $\bar\kappa\approx 6$, and
$A(\bar\kappa)\sim 0.1$, so that $\sim 4\times 10^{11}$ events are
needed.  On our computer system,\footnote{We performed most of our
simulations using GCC on 900 MHz SPARCv9 processors.  Note that this
running time estimate also depends on how much information is collected
during the simulation.}  this requires $\sim 14$ days.

Eq.~(\ref{eqn:runtime}) tells us that the computational cost grows
rapidly as $T'/T$ decreases.  In Simulations~\ref{exp4}, {\ref{exp6}},
and {\ref{exp9}}, where we study the small-$T'$ behavior of the
covariance function, this rapid growth prevents us from taking $T'$ too
small.  However, we do not always need small $T'/T$, and a large
temperature gradient not only reduces the computational cost, it also
takes the system farther out of equilibrium so that some effects are
made more transparent.

The cost also grows rapidly as $N$ increases.  This is why we use short
chains wherever possible, for example in computing the $A$-curve.  In
Simulation~\ref{exp5}, we use $N=8$ because we feel that the amount of
accuracy gained from using longer chains is perhaps not worth the
additional cost.  However, it is not always possible or advisable to use
short chains, particularly where infinite-volume limits are involved.
In such cases, one can sometimes obtain better results by increasing
$T'/T$.

Finally, recall that $A(\bar\kappa)\sim 1/\bar\kappa$ for $\bar\kappa\gg
1$, so the computational cost is $\sim\bar\kappa^3$ for large
$\bar\kappa$.


\subsection*{B\ \ \ Invariant measures}

{\it Proof of Proposition \ref{lemma4.2}:}
The proof of invariance of the measure $\mu^{T,\rho_1} \times \cdots
\times \mu^{T, \rho_N}$ follows closely that of Proposition 3.7 in
{\cite{ey}}.  We refer the reader to {\cite{ey}} for some of the
background notation.  Fix a phase point
$$
\bar z \ = \ (\bar z^{(1)}, \dots, \bar z^{(N)}) \ = \
(\{\bar x^{(1)}_1, \dots, \bar x^{(1)}_{k_1}\}, y^{(1)}; \ \dots; \
\{\bar x^{(N)}_1, \dots, \bar x^{(N)}_{k_N}\}, y^{(N)})~.
$$
We assume $\bar x^{(n)}_1, \dots, \bar x^{(n)}_{k_n}$ are
distinct, and let $\varepsilon, h>0$ be arbitrarily small numbers.
Fix arbitrary $n$ with $1<n<N$. We will compare  
the 3 probabilities, $P_{n, \cdot}$, $P_{n+1, n}$ and $P_{n-1,n}$
defined below:

Let $P_{n,\cdot}$ be the probability that at time $t=0$, in every site
$j$, there are exactly $k_j$ tracers the energies of which lie in
disjoint intervals
$$[\bar x^{(j)}_1, \bar x^{(j)}_1+\varepsilon], \qquad \dots, \qquad
[\bar x^{(j)}_{k_{j}}, \bar x^{(j)}_{k_{j}}+\varepsilon]\ ,
$$
{\it and} before $t=h$, the tracer in site $n$ with
energy in $[\bar x^{(n)}_1, \bar x^{(n)}_1+\varepsilon]$ exits
the site. The number $P_{n+1, n}$ is the probability that at
time $0$, the tracer configuration is as above except that
the tracer with energy in $[\bar x^{(n)}_1, \bar x^{(n)}_1+\varepsilon]$
is in site $n+1$ instead of site $n$ ({\it i.e.}, there are $k_n-1$ tracers
in site $n$ and $k_{n+1}+1$ tracers in site $n+1$), {\it and} before
time $t=h$, this tracer jumps from site $n+1$ to site $n$. The number
$P_{n-1,n}$ is defined analogously with site $n-1$ playing the role
of site $n+1$. 

To prove the invariance of $\mu^{T,\rho_1} \times \cdots \times \mu^{T,
\rho_N}$, three sets of balancing conditions have to be met. The
interaction with tanks is as in [EY]. We verify below the equation
involving interaction between neighbors, namely $P_{n, \cdot} =
P_{n+1,n} + P_{n-1,n}$, and leave the one involving interaction with a
bath to the reader.

Let $\sigma_k, c_k$ and $p_k$ be as in Proposition~\ref{prop2.1}
(Sect.~\ref{sec:ey-equilibrium}).  We will use the shorthand
$\sigma_{k_i} =\sigma_{k_i}(\bar z^{(i)})$, and write $p_k^{(i)}$ to
remind ourselves that $\rho=\rho_i$ at site $i$ ($c_k$ and $\sigma_k$ do
not depend on $\rho$). First,
\begin{eqnarray*}
P_{n,\cdot} & = &
\Pi_{i=1}^N p^{(i)}_{k_i}c_{k_i} \sigma_{k_i} \varepsilon^{k_i+1} \cdot
\sqrt{\bar x^{(n)}_1} e^{\beta \bar x^{(n)}_1} \frac{1}{\varepsilon} \cdot
\int_{\bar x^{(n)}_1}^{\bar x^{(n)}_1+\epsilon } h \sqrt x \frac{1}{\sqrt x}
e^{-\beta x}dx\\
& = & h \cdot \Pi_{i=1}^N p^{(i)}_{k_i}c_{k_i} \sigma_{k_i} 
\varepsilon^{k_i+1} \sqrt{\bar x^{(n)}_1} \ := \ h \cdot Z
\end{eqnarray*}
where $Z$ is defined by the equality above. Next, $P_{n+1,n} = 
\frac{1}{2} I \cdot II \cdot III$ where
\begin{eqnarray*}
I & = & \Pi_{i \neq n, n+1} \ (p^{(i)}_{k_i}c_{k_i} \sigma_{k_i}
\varepsilon^{k_i+1})\ ,\\
II & = & p^{(n)}_{k_n-1} c_{k_n-1} \sigma_{k_n} \sqrt{\bar x^{(n)}_1}
e^{\beta \bar x^{(n)}_1}
\varepsilon^{k_n}\ ,\\
III & = & p^{(n+1)}_{k_{n+1}+1} c_{k_{n+1}+1} \sigma_{k_{n+1}}
\varepsilon^{k_{n+1}+1}
\ \int_{\bar x^{(n)}_1}^{\bar x^{(n)}_1+\varepsilon} h 
\sqrt x \frac{1}{\sqrt x} e^{-\beta x} dx~.
\end{eqnarray*}
This product can be written as
$$
\frac{h}{2} \cdot Z \cdot
\left(\frac{p^{(n)}_{k_n-1}}{p^{(n)}_{k_n}} \frac{c_{k_n-1}}{c_{k_n}}\right)
\cdot \left(\frac{p^{(n+1)}_{k_{n+1}+1}}{p^{(n+1)}_{k_{n+1}}} 
\frac{c_{k_{n+1}+1}}{c_{k_{n+1}}}\right) 
\ = \ \frac{h}{2} \cdot Z \cdot  \left(\frac{T}{2\rho_n}\right) \cdot 
\left(\frac{2\rho_{n+1}}{T}\right)  \ .
$$
The equality above follows from the relation 
$$
\frac{c_kp_k}{c_{k+1}p_{k+1}} = \frac{T}{2\rho}\ ,
$$ which can be derived from the characterization of $\mu^{T,\rho}$
(Proposition~\ref{prop2.1}). An analogous argument holds for
$P_{n-1,n}$, and the desired equality follows from
$$
\frac{1}{2} \left(\frac{\rho_{n+1}}{\rho_n} + \frac{\rho_{n-1}}{\rho_n}
\right) \ = \ 1\ .
$$
\hfill $\square$


\end{document}